\documentclass[12pt]{article}
\pdfoutput=1
\usepackage{a4wide}
\usepackage{color}
\usepackage{amsfonts}
\usepackage{graphicx}
\begin{document}
\rightline{NSF-KITP-09-74}
\rightline{PUPT-2302}
\vspace{2truecm}
\linespread{1.1}

\centerline{\LARGE \bf A general class of holographic superconductors} 
\vspace{0.5cm}

\vspace{1.3truecm}

\centerline{
    {\large \bf S. Franco${}^{a}$,}  
    {\large \bf A. Garc\'{\i}a-Garc\'{\i}a${}^{b,c}$}  
     {\bf and}
    {\large \bf D. Rodr\'{\i}guez-G\'omez${}^{b,d}$}}
    
\vspace{.4cm}
\centerline{{\it ${}^a$ KITP, University of California}} \centerline{{\it Santa Barbara, CA93106-4030, USA}}
\vspace{.4cm}
\centerline{{\it ${}^b$Department of Physics, 
Princeton University}}
\centerline{{\it Princeton, NJ 08544, USA}}
\vspace{.4cm}
\centerline{{\it ${}^c$ The Abdus Salam International Centre for Theoretical Physics}} \centerline{{\it P.O.B. 586, Trieste 34100, Italy}}
\vspace{.4cm}
\centerline{{\it ${}^d$ Center for Research in String Theory, Queen Mary University of London}} \centerline{{\it Mile End Road, London, E1 4NS, UK}}

\vspace{2truecm}

\centerline{\bf ABSTRACT}
\vspace{.5truecm}

\noindent
We introduce a simple generalization of the basic holographic superconductor model in which the spontaneous breaking of a global $U(1)$ symmetry occurs via the St\"uckelberg mechanism. This more general setting allows tuning features such as the order of the transition. The physical vacuum of the condensed phase and the order of the transition are determined by a detailed analysis of the free energy of the system. For first order transitions, we identify a metastable phase above the critical temperature. In this case, the conductivity shows additional poles, thus suggesting that the condensate has internal structure. We comment on the possibility of obtaining second order phase transitions with non mean-field critical exponents.
\newpage
\tableofcontents

\section{Introduction}

The AdS/CFT correspondence \cite{Maldacena:1997re,Gubser:1998bc,Witten:1998qj} is a powerful tool for dealing with strongly coupled CFTs through a weakly coupled dual gravitational description.\footnote{Non-conformal extensions of the correspondence are also known, with \cite{Klebanov:2000hb} the prototypical example.} Conformal field theories are relevant in condensed matter physics near quantum critical points. Hence, the AdS/CFT has the potential to provide a handle on the dynamics of quantum critical points which cannot be attained by other analytical methods (see the excellent reviews \cite{Sachdev:2008ba,Hartnoll:2009sz,Herzog:2009xv} and references therein for a detailed introduction). More generally, close to second order phase transitions, correlation lengths diverge and systems are well approximated by a continuum scale invariant theory. In this paper, we take a more general approach and consider CFTs even when the previous conditions are not met, for example for first order phase transitions. In this case, the CFTs should be regarded as examples of strongly coupled gauge theories with fairly simple gravity duals.

More specifically, we investigate strongly coupled CFT$_{d-1}$'s that exhibit spontaneous symmetry breaking below some critical temperature along the lines of \cite{Gubser:2008px,Hartnoll:2008vx}. Most of the details of this CFT$_{d-1}$, other than the existence of a global $U(1)$ symmetry which is spontaneously broken by the St\"uckelberg mechanism \cite{stu38} are largely irrelevant. We will just assume that the CFT$_{d-1}$ dynamics can give rise to spontaneous symmetry breaking of the $U(1)$ symmetry controlled by an order parameter given by some scalar operator $\mathcal{O}$ of conformal dimension $\Delta_{\mathcal{O}}$ in the CFT$_{d-1}$. Then, as a function of temperature, the system will be either in an ordered (broken symmetry) phase at low temperatures or in a disordered (symmetric) phase at high temperatures. We will assume the CFT$_{d-1}$ has gravity dual, in terms of which we compute the properties of the system. The study of this phase transition, along with the transport properties of the system, is usually beyond standard analytic field theory methods but still accessible by using the AdS/CFT correspondence. 

The central observation of this paper is that, under the very generic premises demanded on the field theory side, namely a CFT$_{d-1}$ with a global $U(1)$ symmetry which can be spontaneously broken, the dual gravitational description can be considerably more general than the set-up considered in  \cite{Gubser:2008px,Hartnoll:2008vx}. Taking an effective field theory approach, we will propose a class of gravity duals with the right ingredients to model the considered CFTs. We expect that all these gravitational duals do indeed yield a phase transition in a similar manner to \cite{Gubser:2008px,Hartnoll:2008vx}.\footnote{Other interesting realizations of holographic superconductors have been introduced in \cite{Gubser:2008zu,Gubser:2008wv,Roberts:2008ns,Basu:2008st,Ammon:2008fc,Herzog:2009ci,Sonner:2009fk,Ammon:2009fe}.} However, the more general setting provided by the St\"uckelberg model for spontaneous symmetry breaking permits tuning certain features of the phase transition, such as its order. Thus our results provide an effective description of different phase transitions in strongly interacting systems induced by the spontaneous breaking of a global $U(1)$ symmetry.\footnote{We are considering that the $U(1)$ symmetry will eventually be gauged. Strictly speaking, a theory with a spontaneously broken global $U(1)$ symmetry corresponds to a superfluid. The holographic description of such system was studied in \cite{Herzog:2008he,Basu:2008bh}.}

The outline of this paper is as follows. In section 2 we introduce a gravity dual of the St\"uckelberg model for spontaneous symmetry breaking 
which depends on a general function $\mathcal{F}$ of the scalar field. We then find the equation of motion of the gravity dual for the simplest choices of $\mathcal{F}$, namely, monomials of degree $n$ of the field.  Finally we provide qualitative arguments, in very much the same spirit as in  \cite{Gubser:2008px}, that a phase transition should occur. In section 3 we prove numerically that the transition indeed exists and that its order depends on $n$. For $n=2$ the transition is second order and  for $n > 2$ it is first order. In both cases we find that there are actually many solutions to the equations of motion. The most stable -and therefore physical- solutions are determined in section 4 by analyzing the free energy. In the case of first order phase transitions we have identified the expected metastable regime around the critical temperature. The transport properties of some of these CFTs are studied in section 5. For first order phase transitions ($n> 2$) we find extra poles in the conductivity. This 
hints the condensate in the ordered phase has internal structure. In section 6, we investigate more general choices of $\mathcal{F}$ and discuss the possibility of non mean-field critical exponents. We end with some conclusions in section 7.

\section{St\"uckelberg model for spontaneous symmetry breaking}

We are interested in strongly coupled CFT$_{d-1}$'s which enjoy a global $U(1)$ symmetry. We suppose that the strong coupling dynamics of the CFT can lead to spontaneous symmetry breaking of such a $U(1)$ symmetry. This symmetry breaking is controlled by some order parameter. We will assume it to be some scalar operator $\mathcal{O}$ of conformal dimension $\Delta_{\mathcal{O}}$ in the CFT$_{d-1}$. Since we assume this CFT$_{d-1}$ has a gravity dual, on very general grounds, the global $U(1)$ symmetry gets mapped to a bulk $U(1)$ gauge symmetry. Likewise, the operator $\mathcal{O}$ translates to a bulk scalar field $\Psi$ with the appropriate mass given by $\Delta_{\mathcal{O}}$. Therefore, the minimal requirements to holographically model the systems of interest is a $U(1)$ gauge field and a scalar; both coupled to gravity and living in an (asymptotically) AdS$_d$ background. Since we are interested in the finite temperature case, the background will actually be a black hole in AdS$_d$. Spontaneous breaking of the boundary $U(1)$ symmetry at some critical temperature translates in the holographic dual into condensation of the scalar. This leads to spontaneously breaking of the bulk gauge symmetry through the Higgs mechanism, which is dual to the spontaneous breaking of the boundary global symmetry \cite{Klebanov:1999tb,Klebanov:2002gr}. 

The condensation of the scalar actually changes the background from the AdS-black hole in the uncondensed phase to a hairy black hole in the condensed phase. The following minimally coupled scalar \cite{Gubser:2008px,Hartnoll:2008vx} is one of the simplest models with this property,

\begin{equation}
S=\int\,\sqrt{g}\,\Big\{ R-\Lambda -\frac{ F^2}{4}-\frac{|D_{\mu}\Psi|^2}{2}-\frac{m^2\,|\Psi|^2}{2}-V(|\Psi|)\Big\},
\label{S_basic_HSC}
\end{equation}
where $D_{\mu}=\partial_{\mu}-i\,A_{\mu}$. We can innocuously re-write this model in a 
St\"uckelberg form by re-writing the charged scalar field $\Psi$ as $\tilde{\Psi}\, e^{ip}$:

\begin{equation}
S=\int\,\sqrt{g}\,\Big\{R - \Lambda -\frac{F^2}{4}-\frac{\partial\tilde{\Psi}^2}{2}-\frac{m^2\,\tilde{\Psi}^2}{2}-\frac{\tilde{\Psi}^2}{2}\big(\partial p-A)^2-V(\tilde{\Psi})\Big\}
\label{rewriten_basic_HSC}
\end{equation}
with $\tilde{\Psi}$ and $p$ real.
The gauge symmetry becomes $A\rightarrow A+\partial \alpha$ together with $p\rightarrow p+\alpha$. So far all we have done is rewriting the model. Nevertheless, it is straightforward to generalize the model preserving gauge invariance. The generalized action reads

\begin{equation}
\label{gmodel}
S=\int\,\sqrt{g}\,\Big\{ -\frac{F^2}{4}-\frac{\partial\tilde{\Psi}^2}{2}-\frac{m^2\,\tilde{\Psi}^2}{2}-\left|\mathcal{F}(\tilde{\Psi})\right|\big(\partial p-A)^2-V(\tilde{\Psi})\Big\}
\end{equation}
where $\mathcal{F}$ is a function of $\tilde{\Psi}$. We take its absolute value to ensure positivity of the kinetic term for $p$.\footnote{One could in principle be worried about the non-analyticity of such an action due to the absolute value. In all the examples we study in the paper, the solutions remain well inside the region of $\mathcal{F}>0$ for all radial positions. We can always think about modifying $\mathcal{F}$ (if necessary) such that the solutions are preserved but $|\mathcal{F}|$ remains non-zero and analytic for all $\Psi$.} We will refer to this model as the {\it St\"uckelberg holographic superconductor}. 
Our generalization of the basic holographic superconductor takes a rather compact form. 

When re-writing (\ref{S_basic_HSC}) in the form (\ref{rewriten_basic_HSC}), it is important to remember that $\tilde{\Psi}$ is the absolute value of a complex field and hence positive definite. Keeping the definition of our model as broad as possible, $\tilde{\Psi}$ in (\ref{gmodel}) takes in general any real value. Nevertheless, as it happens for (\ref{rewriten_basic_HSC}), it is sometimes natural to restrict $\tilde{\Psi}$ to be positive if the model admits a reformulation in terms of a complex field. We will exploit this fact later. The results in \cite{Gubser:2008px,Hartnoll:2008vx} strongly suggest the detailed form of the potential $V$ are largely irrelevant as far as the condensation is concerned. Thus, for the sake of simplicity, we will assume $V=0$ from now on.

\subsection{Choice of $\mathcal{F}$}

In order to determine the action, we need to specify the function $\mathcal{F}$ in (\ref{gmodel}). If $\mathcal{F}$ is an analytic function of 
$\tilde{\Psi}$, it admits a Taylor expansion. In what follows, we will focus mostly on some simple models in which $\mathcal{F}$ is just a monomial of some degree $n$

\begin{equation}
\mathcal{F} \sim \tilde{\Psi}^n.
\label{simple_F}
\end{equation} 
This choice is a good starting point to initiate the analysis of St\"uckelberg superconductors.
In section 6, we will explore more general choices of $\mathcal{F}$.

To be more specific, the theories we consider contain gravity with a negative cosmological constant $\Lambda$ plus a real scalar field $\tilde{\Psi}$, a real pseudoscalar field $p$ and a $U(1)$ gauge field whose coupling constant we call $e$. The action reads,
\begin{eqnarray}
\label{nmodel}
S_n&=&\int d^{d+1}\, \sqrt{g}\,\Big\{ R-\Lambda+\\ \nonumber &&-\frac{F_{MN}F^{MN}}{4e^2}-\frac{\partial_M\tilde{\Psi}\partial^M\tilde{\Psi}}{2}-\frac{m^2}{2}\tilde{\Psi}^2-\Big(\frac{e}{M}\Big)^{n-2}\,\left|\tilde{\Psi}^n\right|\,(\partial_Mp-A_M)(\partial^Mp-A^M)\Big\}\ .
\end{eqnarray}
The scale $M$, required by dimensional analysis, will be set to the unity.\footnote{Eventually these phenomenological models might come from a UV completion such as string theory. It is then natural to assume that $M\sim L^{-1}$, being $L$ the AdS radius. Since we will work in units in which $L=1$, this provides justification to set $M=1$.}

This theory has the aforementioned local gauge invariance under which the $p$ and $A_M$ fields transform as
\begin{equation}
p\rightarrow p+\alpha(x^N)\ ,\qquad A_M\rightarrow A_M+\partial_M\alpha(x^N)\ .
\end{equation}
This symmetry is broken if $\tilde{\Psi}$ develops a VEV. In flat space this would never occur without a Higgs-like potential for $\tilde{\Psi}$. However in a non-trivial background, as in \cite{Gubser:2008px,Hartnoll:2008vx}, the presence of a chemical potential or a charge density might induce the condensation of $\tilde{\Psi}$. Indeed upon a trivial numerical re-scaling, the $n=2$ theory is just the one considered in \cite{Gubser:2008px,Hartnoll:2008vx}, where the existence of a condensed phase was explicitly shown. 

\subsection{Equations of motion}

We are now ready to derive the equations of motion for the different fields in (\ref{nmodel}).  We start with the gravitational field. The Einstein equations are sourced by the matter stress-energy tensor. The latter is given by
\begin{eqnarray}
\frac{T_{MN}}{\sqrt{g}}&=&g_{MN}\,\frac{F_{AB}F^{AB}}{8e^2}+g_{MN}\,\frac{\partial_A\tilde{\Psi}\partial^A\tilde{\Psi}}{4}+g_{MN}\,\frac{m^2}{4}\,\tilde{\Psi}^2+g_{MN}\,\frac{e^{n-2}}{2}\,\tilde{\Psi}^n\,(\partial_Ap-A_A)^2 \nonumber \\ &&+\frac{F_{MB}F^B\,_N}{2e^2}-\frac{\partial_M\tilde{\Psi}\partial_N\tilde{\Psi}}{2}- e^{n-2}\tilde{\Psi}^n\,(\partial_Mp-A_M)(\partial_Np-A_N)\ .
\end{eqnarray}
After a re-scaling $\tilde{\Psi}=e\Psi$,
\begin{eqnarray}
\frac{T_{MN}}{\sqrt{g}}&=&\frac{1}{e^2}\Big(g_{MN}\,\frac{F_{AB}F^{AB}}{8}+g_{MN}\,\frac{\partial_A\Psi\partial^A\Psi}{4}+g_{MN}\,\frac{m^2}{4}\,\Psi^2+g_{MN}\,\frac{\Psi^n}{2}\,(\partial_Ap-A_A)^2 \nonumber \\ &&+\frac{F_{MB}F^B\,_N}{2}-\frac{\partial_M\Psi\partial_N\Psi}{2}- \Psi^n(\partial_Mp-A_M)(\partial_Np-A_N)\Big)\ ;
\end{eqnarray}
In this form it is clear that in the probe limit ($e \to \infty$) gravity is decoupled from the rest of fields. For the sake of simplicity we will work in this limit in the rest of the paper.  

Since there is a negative cosmological constant, we will consider AdS$_{d+1}$ black holes, whose geometry is given by
\begin{equation}
ds^2=-f(r)\,dt^2+\frac{dr^2}{f(r)}+r^2\, d\vec{x}_{d-1}^2\ , \qquad f(r)=\frac{r^2}{L^2}-\frac{M^{d-2}}{r^{d-2}}\ .
\end{equation}
The dynamics of the other fields in this background is described by the action,
\begin{equation}
\label{model}
S_n=\int d^{d+1}\, \sqrt{g}\,\Big\{-\frac{F_{MN}F^{MN}}{4}-\frac{\partial_M\Psi\partial^M\Psi}{2}-\frac{m^2}{2}\Psi^2-\Psi^n\,(\partial_Mp-A_M)(\partial^Mp-A^M)\Big\}\ .
\end{equation}

The equations of motion of these fields are given by the Lagrange equations coming from (\ref{model}) in the above AdS$_{d+1}$ background:
\begin{equation}
\label{eq1}
\partial_A\Big(\sqrt{g}\,F^{AB}\Big)+2\,\sqrt{g}\,\Psi^n\Big(\partial^Bp-A^B\Big)=0\ ,
\end{equation}
\begin{equation}
\label{eq2}
\partial_B\Big(\sqrt{g}\,\Psi^n\,\Big(\partial^Bp-A^B\Big)\Big)=0\ ,
\end{equation}
and
\begin{equation}
\label{eq3}
\partial_M\Big(\sqrt{g}\,\partial^M\Psi\Big)-m^2\,\sqrt{g}\,\Psi-n\,\sqrt{g}\,\Psi^{n-1}\,\Big(\partial_Mp-A_M\Big)^2=0\ .
\end{equation}
We just keep (\ref{eq1}) and (\ref{eq3}) since (\ref{eq2}) can be obtained by acting with $\partial_B$ on (\ref{eq1}). 

We use the gauge freedom to fix $p=0$. In addition, we consider that only the time component $A_0$ of the gauge field, which we will call $\Phi$, is turned on. Since we are interested in hair-like solutions we will also assume that both $\Psi$ and $\Phi$ are only functions of $r$. Then, particularizing for the background of interest, the relevant equations become,
\begin{equation}
\Phi''+\frac{d-1}{r}\,\Phi'-2\,\frac{\Psi^n}{f}\,\Phi=0\ ;
\end{equation}
and
\begin{equation}
\Psi''+\Big(\frac{d-1}{r}+\frac{f'}{f}\Big)\,\Psi'-\frac{m^2}{f}\,\Psi+n\,\frac{\Psi^{n-1}}{f^2}\,\Phi^2=0\ ;
\end{equation}
where the prime denotes derivative with respect to $r$.

For $n>1$ it is clear that a solution of the above equations with the required boundary conditions is $\Psi=0$ and $\Phi=\mu-\frac{\rho\, r_H^{d-2}}{r^{d-2}}$. This is just the normal (symmetric) phase. We will later show that for sufficiently low temperatures ($\sim r_H$), a condensed solution representing the symmetry breaking phase exists.

It is convenient to define the dimensionless coordinate $z=\frac{r_H}{r}$, where $r_H$ is the horizon radius
\begin{equation}
r_H^d=L^2M^{d-2}\ .
\end{equation}
In these coordinates the horizon sits at $z=1$ while the boundary of the AdS space is at $z=0$. 
The equations of motion are given by,
\begin{equation}
\label{aa}
\Phi''+\Phi'\,\frac{3-d}{z}-\frac{2\,L^2\Psi^n}{z^2\,(1-z^d)}\,\Phi=0\ ,
\end{equation}
and
\begin{equation}
\label{bb}
\Psi''+\Psi'\,\frac{(1-d-z^d)}{z\,(1-z^d)}-\frac{m^2}{z^2\,(1-z^d)}\,\Psi+\frac{n}{r_H^2}\,\frac{\Phi^2}{(1-z^d)^2}\,\Psi^{n-1}=0\ ;
\end{equation}
where now prime denotes derivative with respect to $z$.

\subsection{Boundary conditions}

In order to solve the equations above, we have to fix the behavior of the field at the horizon and the boundary.
Close to the boundary the scalar field has free field behavior. Therefore, 
\begin{equation}
\label{cc}
\Psi=\Psi_+\,r_H^{\lambda_+}\,z^{\lambda_+}+\Psi_-\,r_H^{\lambda_-}\,z^{\lambda_-}\ ,\qquad \lambda_{\pm}=\frac{d\pm\sqrt{d^2+4\,m^2\,L^2}}{2}\ .
\end{equation}
Since we are interested in spontaneous symmetry breaking, we focus on normalizable modes, such that their coefficient is proportional to the VEV of the dual operator $\mathcal{O}_{\pm}$ at the boundary. 
In order to interpret the solution to (\ref{aa}) and (\ref{bb}) as hair for the black hole the fields must remain regular everywhere. Normalizability of $\Phi\,dt$ requires that $\Phi$ vanishes at the horizon. Furthermore, regularity of the solution requires $\Psi$ to be finite at the horizon
\begin{equation}
\Psi=a +b \,(1-z)+\cdots
\end{equation}
 and to decay as (\ref{cc}) at the boundary.

Likewise, close to the boundary, $\Phi$ behaves as
\begin{equation}
\Phi \sim \mu-\frac{\rho}{r_H^{d-2}}\,z^{d-2}\ ,
\end{equation}
where $\mu$ is a chemical potential and $\rho$ a charge density. We will later solve the equations numerically and clarify whether our model has an ordered phase for sufficiently low temperature. 
In the next section we provide heuristic arguments that this is the case.
\subsection{Qualitative arguments for the existence of the transition}
We note that the equation of motion for $\Psi$ is the same as the one we would obtain from an effective Lagrangian
\begin{equation}
S_{eff}=\int\,\sqrt{g}\,\Big(\partial_M\Psi\partial^M\Psi+m^2\Psi^2+\lambda_{eff}\,\Psi^n\Big)\ ,
\end{equation}
where the effective coupling is
\begin{equation}
\label{pot}
\lambda_{eff}=-\frac{z^2\,\Phi^2}{r_H^2\,(1-z^d)}\ .
\end{equation}
Thus, we have effectively a theory for a scalar field with a given potential of order $n$ (let us assume $n>1$). A crucial observation is that this coupling is negative. These types of unbounded potentials are known to give rise to hairy black holes (see e.g. \cite{Torii:2001pg,Gubser:2005ih,Hertog:2006rr}). This is our first indication that a condensed phase might actually exist for any $n$. Nevertheless, this is not sufficient to argue that a phase transition indeed exists. 
In this situation, the hair is in some sense more a property of the boundary than of the horizon. However, this is not the case for the potential above, since the coupling (\ref{pot}) depends on properties of the horizon through $\Phi^2$ (which depends on the chemical potential and the density), $r_H$ and the extra redshift factors. Then, it is not surprising that, as we will later see, the potential is sufficiently strong to trigger the existence of hair, at least within a certain window of parameters.

Let us now consider the $\Psi$ equation of motion, which can be rewritten as,
\begin{equation}
\Psi''+\Psi'\,\frac{(1-d-z^d)}{z\,(1-z^d)}-\frac{\Psi}{z^2\,(1-z^d)}\,\Big(m^2+\frac{n}{r_H^2}\,\frac{z^2\,\Phi^2}{(1-z^d)}\,\Psi^{n-2}\Big)=0.\
\label{eom3}
\end{equation}
Close to the boundary, the potential-like term is suppressed with respect to the mass term. This explains why the existence of a condensed phase is not a priori guaranteed, since the potential effectively does not extend all the way to the boundary. On the contrary, any non-vanishing expectation value for the scalar field at the horizon would lead to an instability in the electric field, in the same spirit of \cite{Gubser:2008px,Hartnoll:2008vx}.

The value of $n$ does indeed affect the details of the condensation. The reason is that $\Psi$ goes from $\Psi_0$ in the horizon to $z^{\#}$ at the boundary. Since $\Psi$ appears raised to the power $n-2$ in (\ref{eom3}), it is natural to expect that the larger $n$ is, the more sudden this contribution kicks in. From this heuristic argument, we expect that the phase transition becomes sharper as $n$ increases (in fact, we will later see that it becomes first order for $n >2$). Another consequence of this argument is that the critical temperatures should decrease for larger $n$. 

\section{Numerical evidence for a phase transition}

In this section we use the shooting method to numerically solve the equations of motion (\ref{eom1}) and (\ref{eom2}) with the boundary conditions discussed in the previous section. In order to compare with \cite{Hartnoll:2008vx}, we work in the canonical ensemble. In this ensemble, the density $\rho$ is kept fixed and the chemical potential is determined by the solution of the equations of motion. Moreover, for concreteness, we assume $d=3$ and $mL^2=-2$. We do not expect that other choices will qualitatively modify our results. 
For the above values of $d$ and $m$ the behavior of the scalar field at the boundary is
\begin{equation}
\Psi=\frac{\Psi_1}{r_H}\, z+\frac{\Psi_2}{r_H^2}\, z^2\ .
\end{equation}
We note that the mass, though tachyonic, is above the Breitenlohner-Freedman (BF) bound in AdS$_4$.\footnote{In fact, as discussed in \cite{Horowitz:2008bn}, the mass term plays no important role in the existence of condensation for the scalar. The value of the mass affects various characteristics of the resulting superconductor, though.} In addition, both modes $\Psi_1$ and $\Psi_2$ are in this case normalizable. We define $\langle\mathcal{O}_i\rangle=\Psi_i$. In order to have a well defined solution, we have to choose a quantization in which only one of the  $\langle\mathcal{O}_i\rangle$ is non-vanishing. 
The analysis of the free energy in the next section will confirm this point.

For the numerical calculation we absorb $r_H$ in $\Phi$ as $\varphi=\frac{\Phi}{r_H}$ and define $\chi=\frac{\Psi}{z}$. This choice is numerically more efficient since the equations do not depend explicitly on the temperature and the $\Psi$ expansion at the boundary starts with a constant instead of with a $z$ suppression. In terms of these new fields, the equations of motion read
\begin{equation}
\label{eom1}
\varphi''-\frac{2\,L^2\,z^{n-2}}{(1-z^3)}\,\chi^n\,\varphi=0\ ,
\end{equation}
and
\begin{equation}
\label{eom2}
\chi''-\frac{3z^3}{z\,(1-z^3)}\,\chi'-\frac{z}{(1-z^3)}\,\chi+\frac{n\, z^{n-2}\, \varphi^2}{(1-z^3)^2}\,\chi^{n-1}=0\ .
\end{equation}
 
In what follows, we focus on the cases $n=2$ and $3$. We have also investigated $n > 3$ and observed that the main qualitative features are not changed. 

The condensed phase, where $\langle\mathcal{O}_i\rangle\ne 0$, only exists below some temperature, which we denote $T_0$. The critical temperature $T_c$, is the one at which the free energy of the condensed phase becomes smaller than the one for the uncondensed phase. Generically, $T_c \leq T_0$. For second order phase transitions, $T_c=T_0$. These issues are discussed in detail in section \ref{section_free_energy}.

\begin{figure}[h!]
\centering
\includegraphics[scale=0.7]{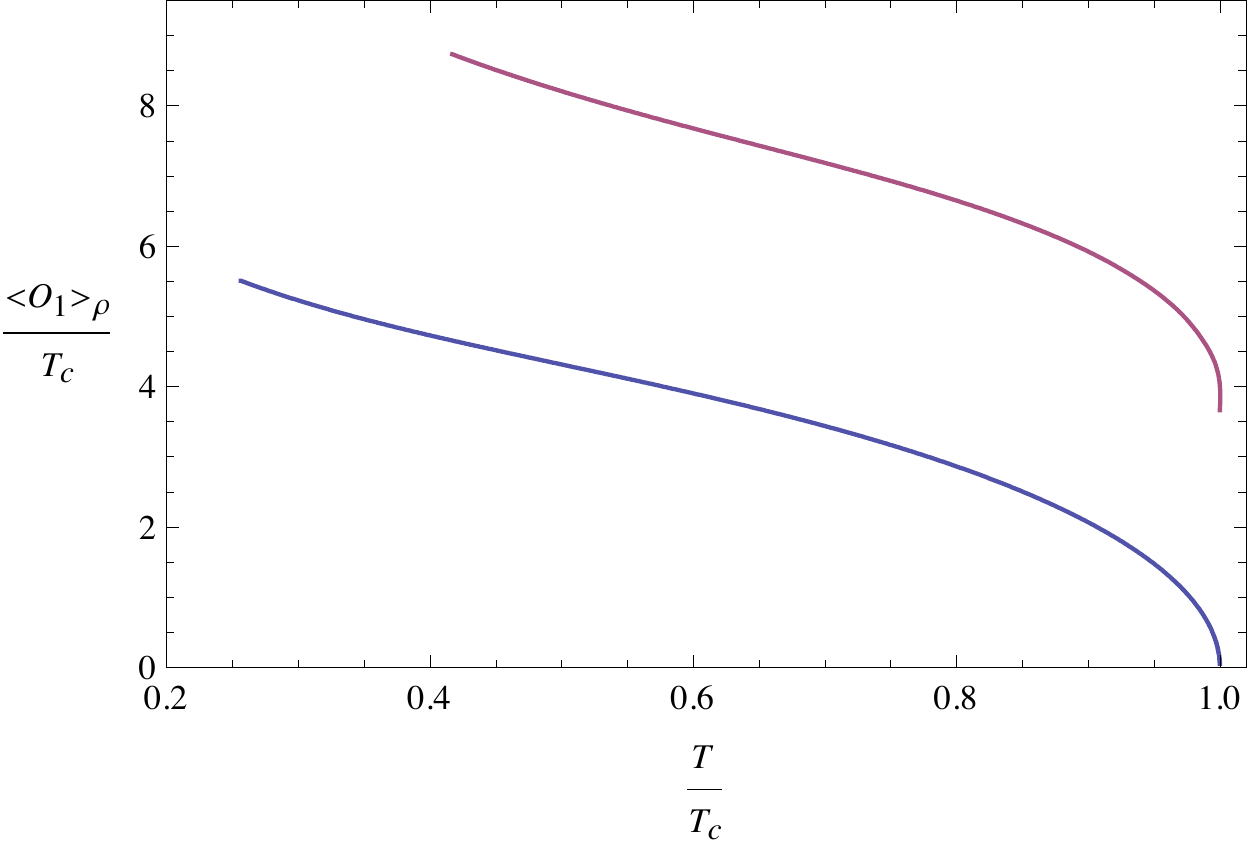}\caption{Normalized condensate $\langle\mathcal{O}_1\rangle$ as a function of temperature at fixed $\rho=-1$. The blue and magenta curves correspond to $n=2$ and $n=3$, respectively. For $n=2$, $T_c \sim 0.268$ while for the $n=3$, $T_c\sim 0.187$.}
\label{O1CanonicalEnsemble}
\end{figure}

\begin{figure}[h!]
\centering
\includegraphics[scale=0.7]{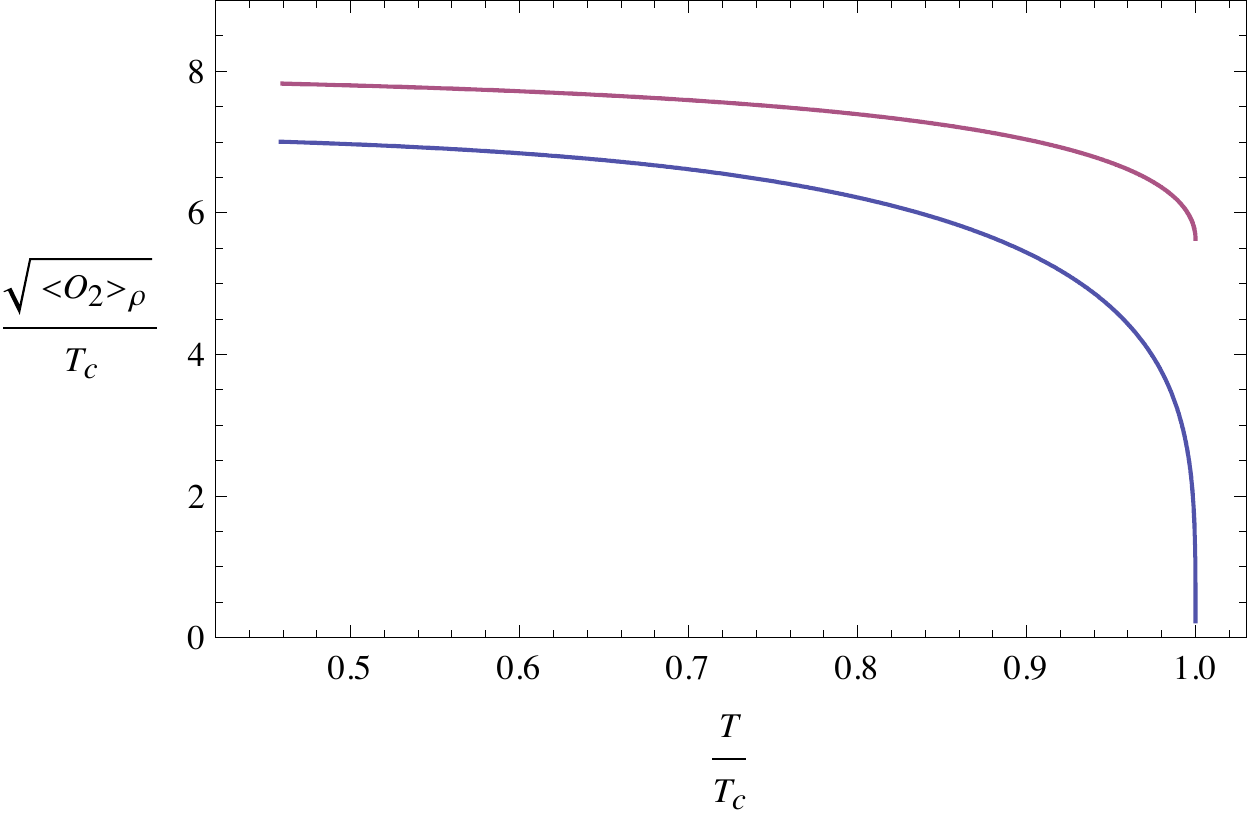}\caption{Normalized condensate $\sqrt{\langle\mathcal{O}_2\rangle}$ as a function of temperature at fixed $\rho=-1$. The blue and magenta curves correspond to $n=2$ and $n=3$, respectively. For $n=2$, $T_c \sim 0.141$; while for $n=3$, $T_c\sim 0.113$.  }
\label{O2CanonicalEnsemble}
\end{figure}

In figure \ref{O1CanonicalEnsemble} we present results for the $\mathcal{O}_1$ condensate normalized by the critical temperature for $n= 2,3$. 
%
Figure \ref{O2CanonicalEnsemble} shows a similar plot for the normalized $\mathcal{O}_2$ condensate. Similarly to what happens for $n=2$, it is possible to derive the $n=3$ model from one with a complex scalar such that $\Psi$ is constrained to be positive.\footnote{If we allow $\Psi$ to take any real value, both the $n=2$ and $3$ models have a $\Psi \to -\Psi$ symmetry that allows us to focus only on positive values.} 

\begin{figure}[h!]
\centering
\includegraphics[scale=0.7]{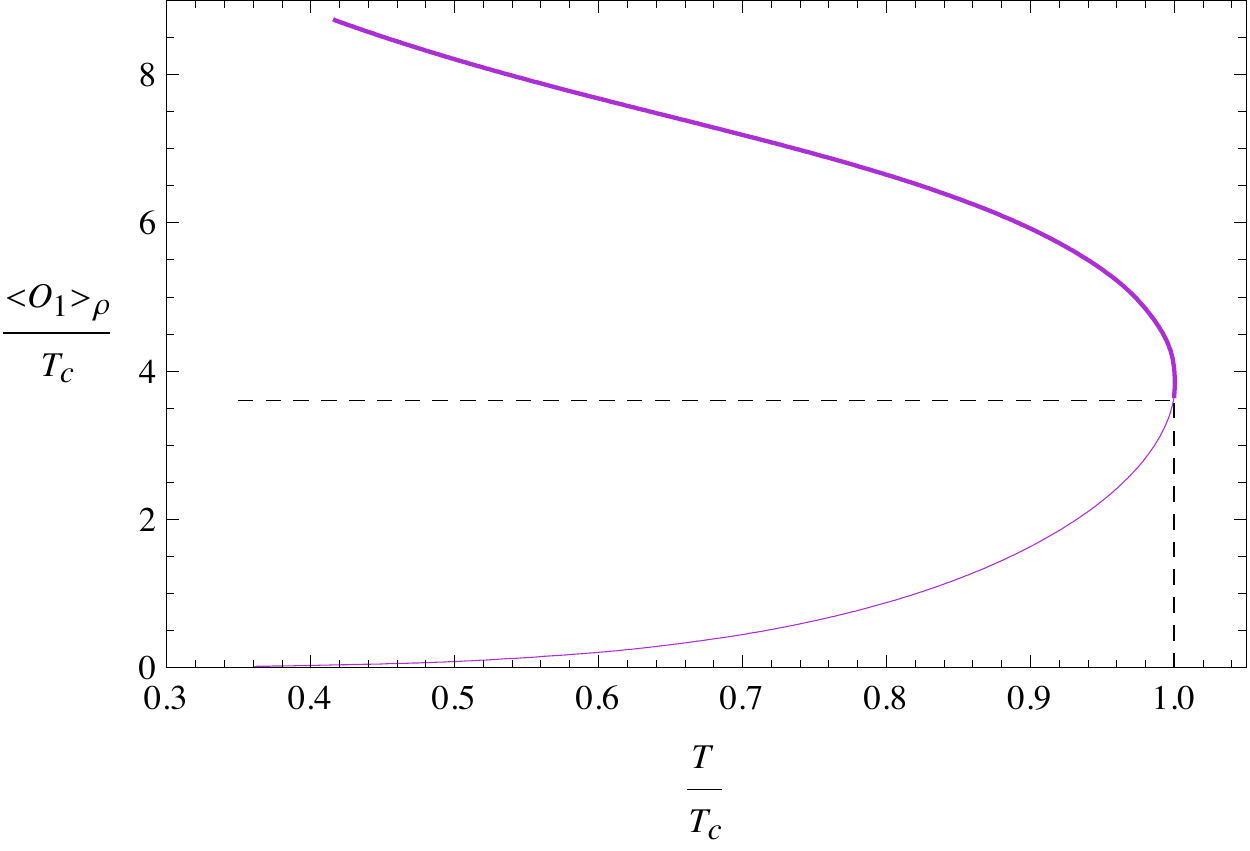}\caption{Normalized condensate $\langle\mathcal{O}_1\rangle$ as a function of temperature at fixed $\rho=-1$ for $n=3$. The thicker curve indicates the physical piece of the full solution.}
\label{O1fullCanonicalEnsemble}
\end{figure}

Figures \ref{O1CanonicalEnsemble} and \ref{O2CanonicalEnsemble} indicate that the transition is first order for $n=3$ (preliminary analysis indicate that the same behavior occurs for $n > 3$). In other words, the condensate remains finite at the temperature $T_c$ at which the free energies of the condensed and uncondensed phases become equal. In section 4, we investigate this issue in greater detail. Indeed, in figures \ref{O1CanonicalEnsemble} and \ref{O2CanonicalEnsemble} the $n=3$ curve is the upper half of the full solution. In figure \ref{O1fullCanonicalEnsemble}, we show the full solution for the $\mathcal{O}_1$ condensate (a very similar plot holds for $\mathcal{O}_2$). Figure \ref{O1fullCanonicalEnsemble} shows that there are two possible values of the condensate for each temperature.\footnote{ A similar behavior was first observed in \cite{Herzog:2008he}, in the case of holographic superfluids in the presence of a non-zero fluid velocity.} Since the upper half of the curve has the largest condensate, it is natural to assume (we confirm this in section 4 by studying the free energy) that the curves in figures \ref{O1CanonicalEnsemble} and \ref{O2CanonicalEnsemble} correspond to the physical solutions.

Finally we note that according to figures \ref{O1CanonicalEnsemble} and 
\ref{O2CanonicalEnsemble} (and similar ones for $n>3$, that we do not exhibit here)  the value of the condensate increases with $n$. This suggests that the attractive interaction that leads to the formation of the condensate becomes stronger as $n$ is increased. 

\subsection{The $n=2$ model: a phase space for the system}

In this section, we investigate in more detail the solutions to (\ref{eom1}) and (\ref{eom2}) for $n=2$ and either $\mathcal{O}_1 $ or 
$\mathcal{O}_2$ equals to zero. We will see that, for sufficiently low temperatures, these equations have multiple solutions or ``branches".\footnote{The analysis of the free energy of these solutions (see section \ref{section_free_energy}) indicates the system condenses into the branch that shows up at the highest temperature, which also corresponds to the one with the largest condensate. } An interesting way of visualizing multiple branches is to momentarily allow $\mathcal{O}_i\neq 0$ and search for solutions to (\ref{eom1}) and (\ref{eom2}), still at fixed $\rho$. As we will see in the next section, one of the condensates must vanish in order to have a critical point of the free energy, which is a necessary condition for the stability of the solution. Thus, out of the resulting curve in the $(\mathcal{O}_1,\mathcal{O}_2)$ plane, we can focus just on the intersections with the axes, at which one of the $\mathcal{O}_i$ vanishes. We will later determine which of these branches is indeed stable, i.e. a minimum of the free energy. In some sense, the $(\mathcal{O}_1,\mathcal{O}_2)$ plane can be regarded as a sort of phase space for the system, which contains interesting information.

As we said, we now focus on $n=2$. Given that $T_c$ for $\mathcal{O}_1$ is bigger than $T_c$ for $\mathcal{O}_2$, we can identify three different regions depending on the temperature:

\bigskip

\begin{center}
\begin{tabular}{ccccc}
Region I: $T_c^{\mathcal{O}_1}<T$ & \ \ \ & Region II: $T_c^{\mathcal{O}_2}<T<T_c^{\mathcal{O}_1}$& \ \ \ & Region III: $T<T_c^{\mathcal{O}_2}$
\end{tabular}
\end{center}

\subsubsection*{Region I}

In this region, $T$ is bigger than any of the two critical temperatures. Therefore, we should expect no cuts with the axes. The resulting plot, which satisfies these expectations, can be seen on the left of figure \ref{spiral_Regions_I_II}. This behavior indicates that the curves in figures 1 and 2 are the first ones to appear as we decrease the temperature and hence determine the critical temperature $T_c$. For $T > T_c$ the system is in the non-condensed phase.

\bigskip

\subsubsection*{Region II}

\begin{figure}[h!]
\centering
\includegraphics[scale=0.6]{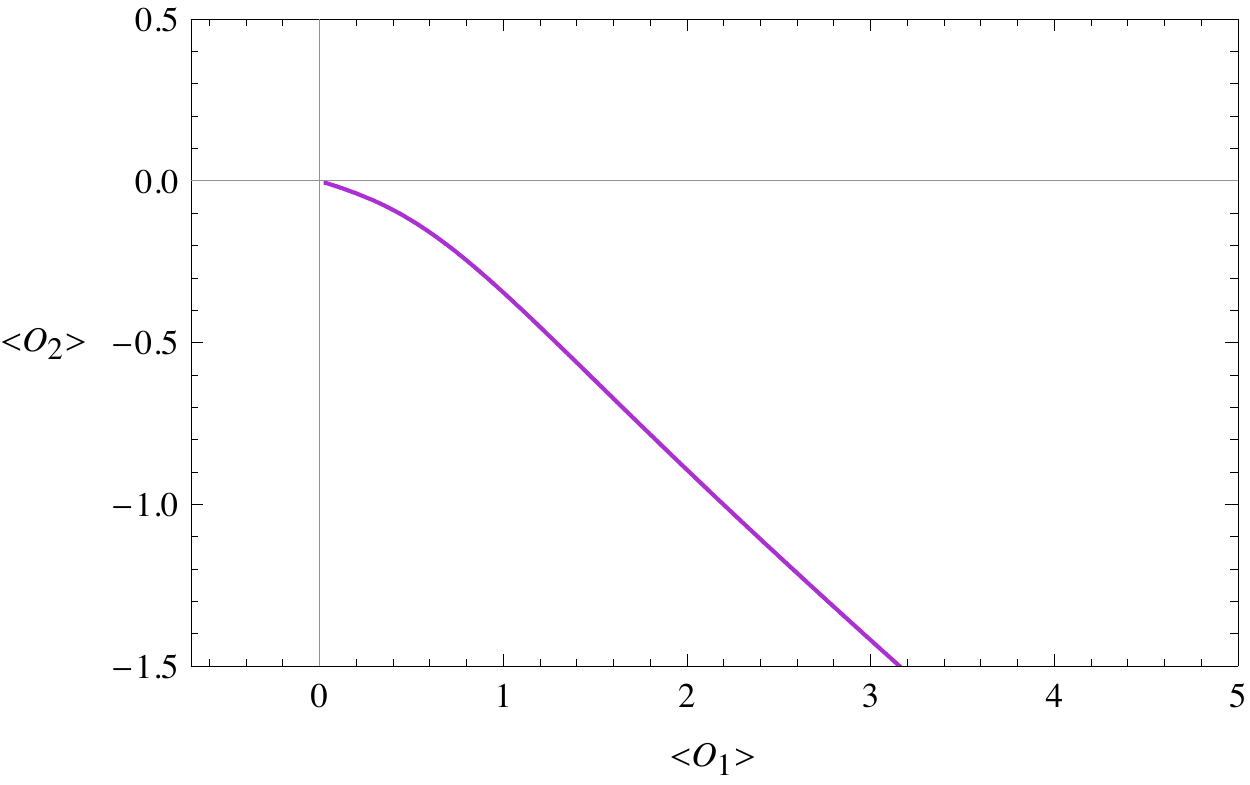} \hspace{.2cm}
\includegraphics[scale=0.6]{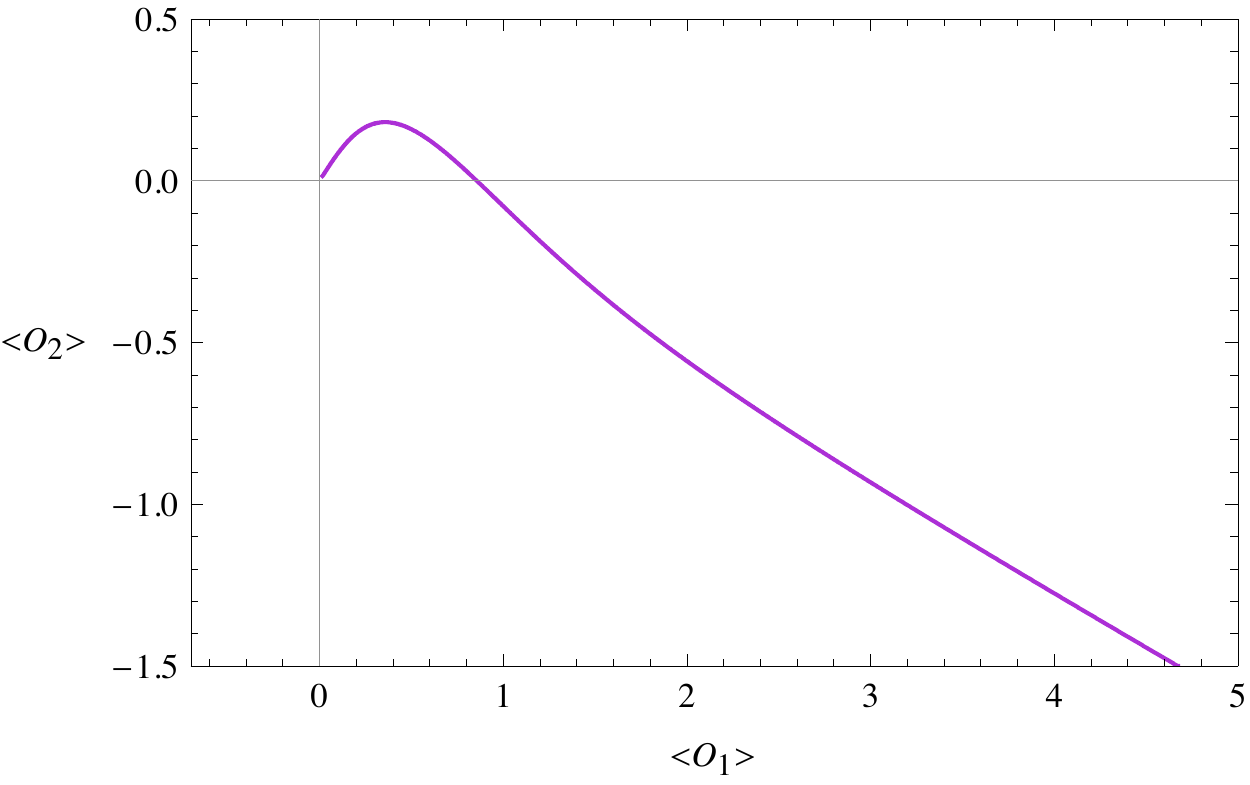} \\
\vspace{-.1cm}\hspace{1.1cm}{\footnotesize $T=0.30$} \hspace{6.5cm} {\footnotesize $T=0.20$} 
\caption{$(\mathcal{O}_1,\mathcal{O}_2)$ plane at $T=0.30$ (Region I) and at $T=0.20$ (Region II).}
\label{spiral_Regions_I_II}
\end{figure}

In this region, $T$ is smaller than the critical temperature for $\mathcal{O}_1$, but larger than $T_c$ for $\mathcal{O}_2$. We therefore expect a cut with the horizontal axis but no cut with the vertical axis, as shown on the right of figure \ref{spiral_Regions_I_II}.

\bigskip

\subsubsection*{Region III}

\begin{figure}[h!]
\centering
\includegraphics[scale=0.5]{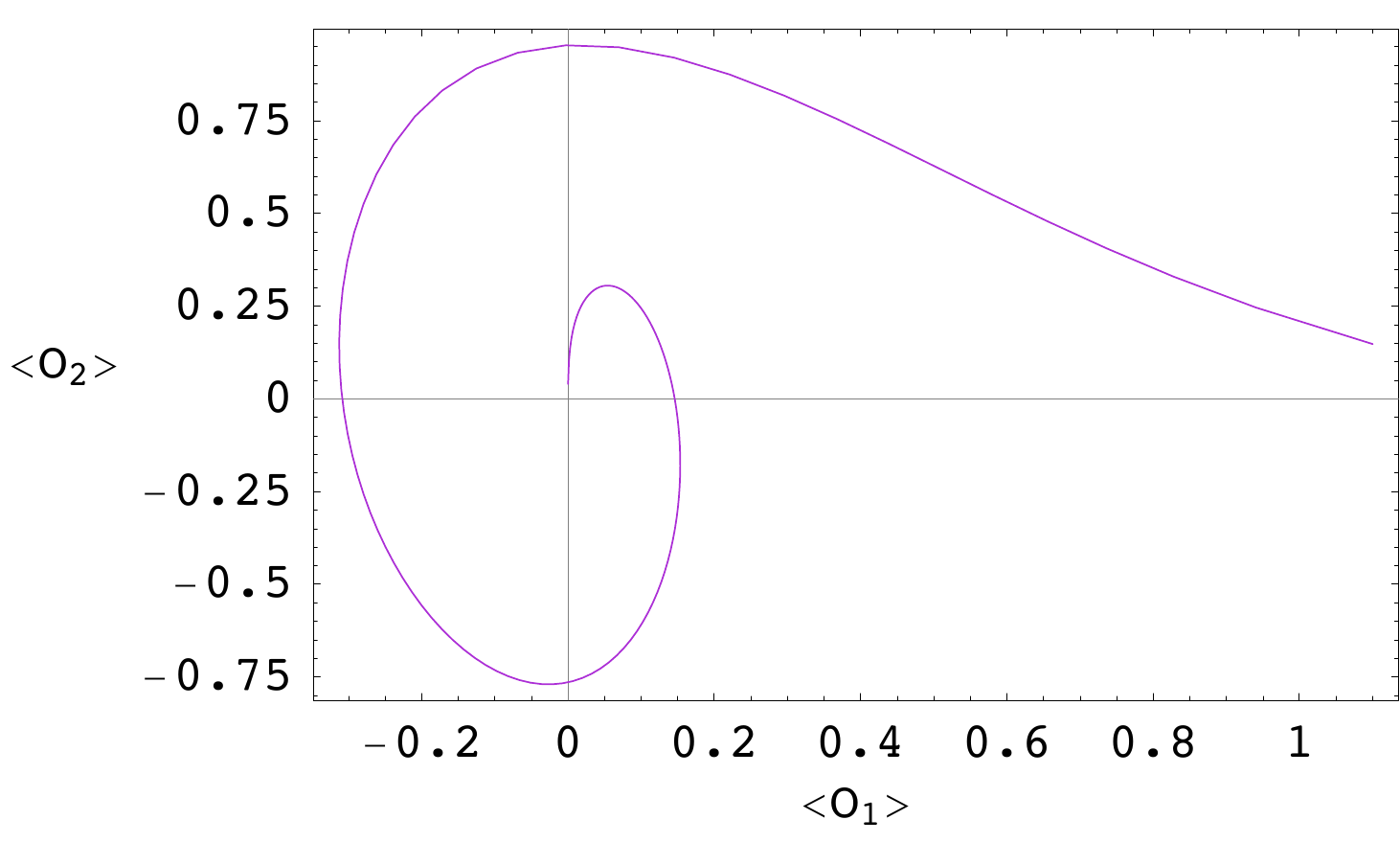} \hspace{.2cm}
\includegraphics[scale=0.5]{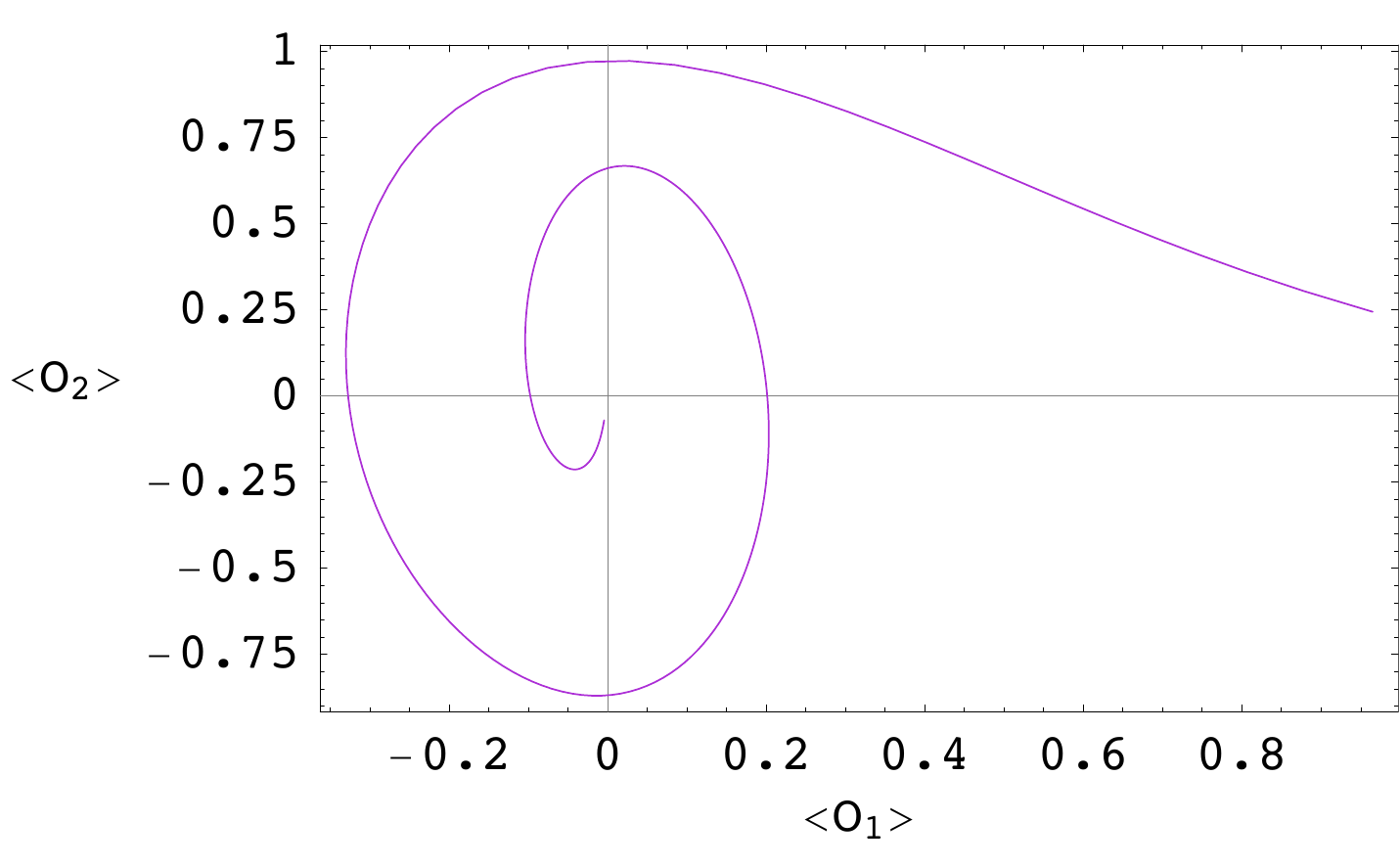} \\
\vspace{-.3cm}\hspace{1.7cm}{\footnotesize $T=0.075$} \hspace{6cm} {\footnotesize $T=0.065$} \\
\vspace{.4cm}\includegraphics[scale=0.5]{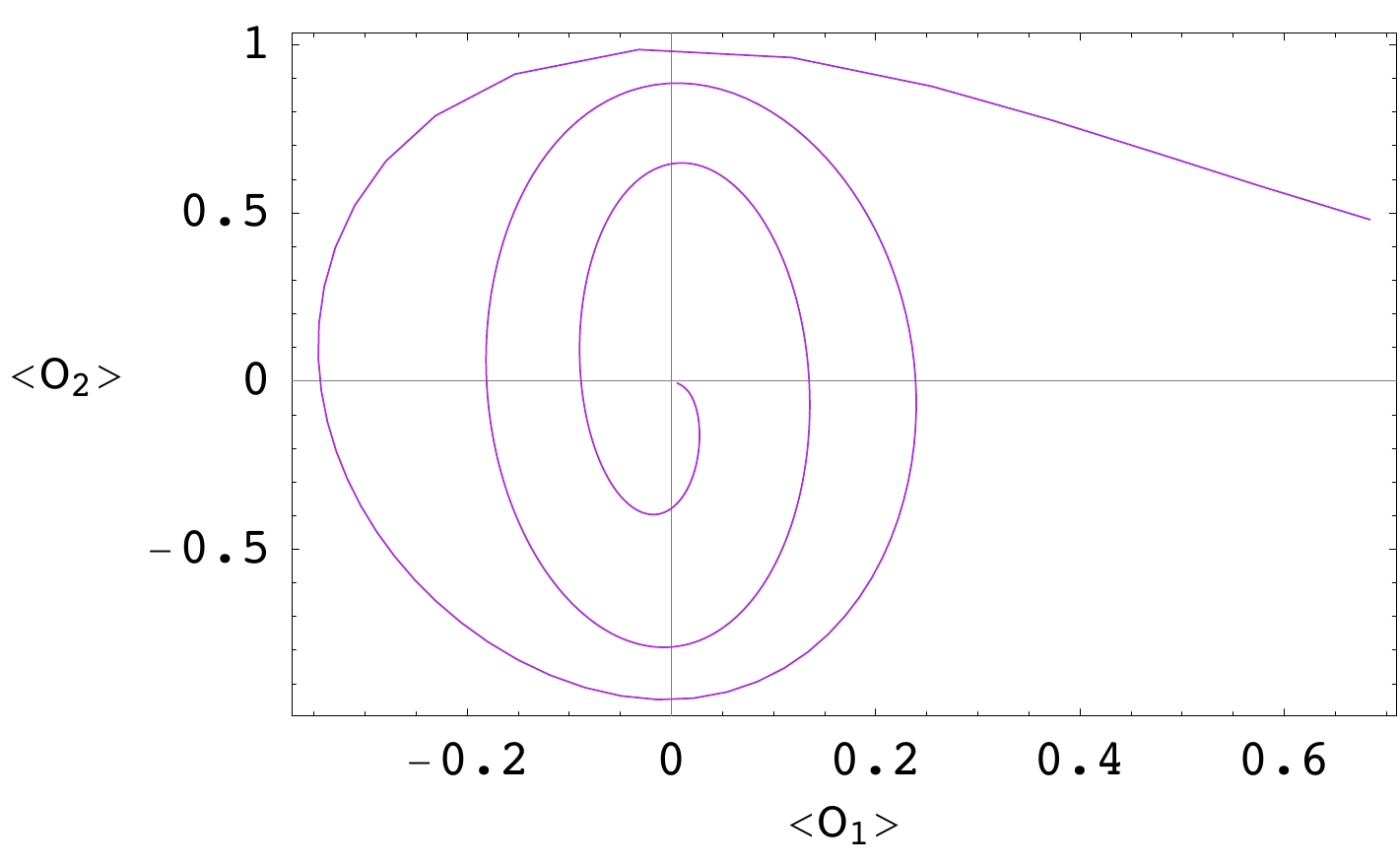} \\
\vspace{-.2cm}\hspace{1.4cm}{\footnotesize $T=0.050$}
\caption{$(\mathcal{O}_1,\mathcal{O}_2)$ plane for $T=0.075$, $0.065$ and $0.050$, well inside Region III.}
\label{spirals_Region_III}
\end{figure}

In this region, $T$ is smaller than both $T_c^{\mathcal{O}_2}$ and $T_c^{\mathcal{O}_1}$. We then expect cuts with both axes. In fact, for sufficiently low temperatures, the spiral circles the origin multiple times. As the temperature is decreased, new turns emanate from the origin. As a result, for very low temperatures, there are multiple intersections with the axes (i.e. branches). Figure \ref{spirals_Region_III} shows spirals for different temperatures. 

\begin{figure}[h!]
\centering
\includegraphics[scale=0.7]{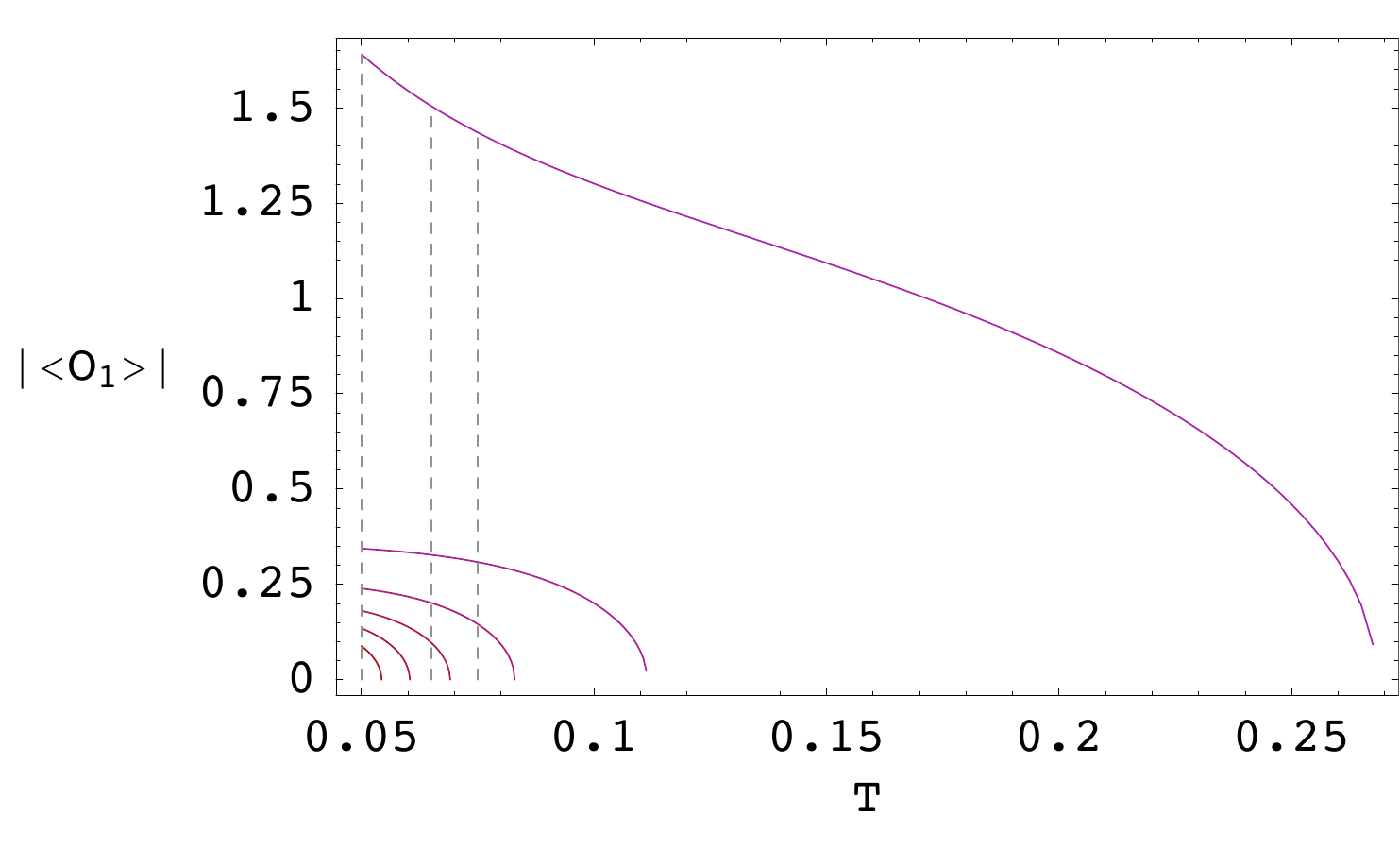}
\caption{$|\mathcal{O}_1|$ as a function of the temperature for the different branches. Notice that, in order to simplify the comparison with figure \ref{spirals_Region_III} we have not normalized the condensate by the temperature as in previous plots. The dashed lines indicate the three temperatures considered in figure \ref{spirals_Region_III}.}
\label{branches}
\end{figure}

Figure \ref{branches} shows the corresponding $\mathcal{O}_2=0$ branches as a function of temperature. In section 4, we will show that the new branches have larger free energy than the ones considered in figures 1 and 2, corresponding to the outermost cuts. Furthermore, while all the branches are critical points of the free energy (i.e. its derivative with respect to the vanishing $\mathcal{O}_i$ is zero), only the outermost branches correspond to 
local minima.

As we have explained, the outermost solution corresponds to the one with the largest $T_c$. This suggests that it should be the one preferred by the system. This picture will be confirmed by the study of the system free energy.

It would be interesting to perform the same analysis for general choices of the function $\mathcal{F}$ and explore whether a behavior similar to the one of the $n=2$ model holds in other examples.

\section{The free energy}

\label{section_free_energy}

In this section we analyze the free energy for $n=2 ,3$. We first review how to compute the free energy in models with gravity duals. 

\subsection{General considerations}

The free energy of a system is given by its on-shell action. In the models we are interested in, this simple procedure involves a few subtleties that we address following \cite{Herzog:2008he}.

Let us start with a little digression about theories in spaces with boundary. In this case, there is a boundary term in addition to the equations of motion in the bulk. So far we have not been careful about this contribution, assuming that the necessary boundary counter-terms were suitably added. To have a well-defined variational problem, not only the bulk equations of motion must be satisfied, but also the boundary term must vanish. Usually, for a theory in a space without boundary, this is achieved by the physically reasonable demand that fields vanish at infinity. However if the space where the theory lives has a boundary we have to tackle the interplay of the boundary conditions and the boundary term. In our case, the boundary term is

\begin{equation}
\label{delta}
\int_{Boundary}\,\Big[\sqrt{g}\,\Big( F_{rt}\,\delta \Phi-g^{rr}\,\partial_r\Psi\, \delta\Psi\Big)\Big]_{Boundary}
\end{equation}

We can take $\delta\Phi_{Boundary}=0$. Since $\delta\Phi=\delta\mu$ at the boundary, this choice corresponds to the grand canonical ensemble, in which $\mu$ is fixed. If, in addition, we choose boundary conditions such that $\delta\Psi_1=0$, the second term also vanishes. The boundary value of $\Psi$ is fixed and, consequently, it corresponds to the choice $\mathcal{O}_1$ quantization.

Under these assumptions we have a well-defined variational problem. Nevertheless, the action is still divergent. In order to regularize it, we introduce a cut-off $r_B$ before the boundary ($r \to \infty$), which we will eventually send to infinity. One can verify that

\begin{equation}
S_{On\,\,Shell}^{(1)}=\frac{\mu\rho}{2}+\frac{r_B \Psi_1^2}{2\,L}+\frac{3\,\Psi_1\,\Psi_2}{2\, L}-\frac{n\, L^2\, r_H}{2}\int_0^1dz\, \frac{\Psi^n\Phi^2}{z^2\, (1-z^3)}
\end{equation}
where the superscript $(1)$ stands for the choice of $\mathcal{O}_1$ quantization.

In order to carry out a holographic renormalization of the free energy, we must add the boundary counter-term,

\begin{equation}
(\Delta S_{Boundary})^{(1)} =-\frac{1}{2}\,\int_{Boundary}\, \left[\sqrt{\gamma}\, \Psi^2\right]_{Boundary}
\end{equation}
where $\gamma$ is the induced metric at the boundary.

The renormalized free energy $W^{(1)}=-S_{On\,\,Shell}^{(1)}-(\Delta S_{Boundary})^{(1)}$ is given by

\begin{equation}
\label{W1}
W^{(1)}=-\frac{\mu\, \rho}{2}-\frac{\mathcal{O}_1\, \mathcal{O}_2}{2}+\frac{n\,L^2\, r_H^3}{2}\,\int_0^1dz\, \frac{z^{n-2}\,\chi^n\,\varphi^2}{1-z^3}.
\end{equation}
Once again, the superscript indicates that we have chosen the $\mathcal{O}_1$ quantization.

As explained in \cite{Klebanov:1999tb}, $\mathcal{O}_1$ and $\mathcal{O}_2$ are canonically conjugated variables. We can see this by fixing  $\delta \Psi_2$ rather than $\delta\Psi_1$ in our general variational problem. In order to do this, we note that, in the grand canonical ensemble, the boundary action (\ref{delta}) reads $ [-r_B^4\,\partial_r\Psi\, \delta{\Psi}]_{Boundary}$. Thus, it can be cancelled by the variation of
\begin{equation}
S_{Boundary}=\int_{Boundary}\, \Big[r^4\, \Psi\, \partial_r\Psi\Big]_{Boundary}.
\end{equation}
Therefore, by considering $S_{Bulk}+S_{Boundary}$, we have a well-defined variational problem provided $\Psi_2$ is fixed at the boundary. 
It is easy to check that the on-shell action is again divergent. As before, it can be renormalized by adding the appropriate counterterm $(\Delta S)^{(2)}=-(\Delta S)^{(1)}$, resulting in

\begin{equation}
W^{(2)}=-\frac{\mu\, \rho}{2}+\frac{\mathcal{O}_1\, \mathcal{O}_2}{2}+\frac{n\,L^2\, r_H^3}{2}\,\int_0^1dz\, \frac{z^{n-2}\,\chi^n\,\varphi^2}{1-z^3}.
\end{equation}
Then, we can think about $W^{(2)}$ as the Legendre transformation of $W^{(1)}$. In particular,

\begin{equation}
\frac{\partial W^{(1)}}{\partial\mathcal{O}_1}=-\mathcal{O}_2\qquad \frac{\partial W^{(2)}}{\partial\mathcal{O}_2}=\mathcal{O}_1.
\end{equation}
Thus, the local extrema of $W^{(1)}$ sit at vanishing $\mathcal{O}_2$ and, conversely, the extrema of $W^{(2)}$ are at $\mathcal{O}_1=0$. This justifies our previous assumption that only one of the $\Psi_i$ is non-vanishing for physical solutions.

Similarly we can keep $\rho$, instead of $\mu$, fixed at the boundary, i.e. work in the canonical ensemble \cite{Herzog:2008he}. In what follows, we focus on the grand canonical ensemble with $\mu=-1$. Moreover, we mostly present results for the $\mathcal{O}_1$ quantization.

\subsection{Free energy for the $n=2$ model}

\begin{figure}[h!]
\centering
\includegraphics[scale=0.5]{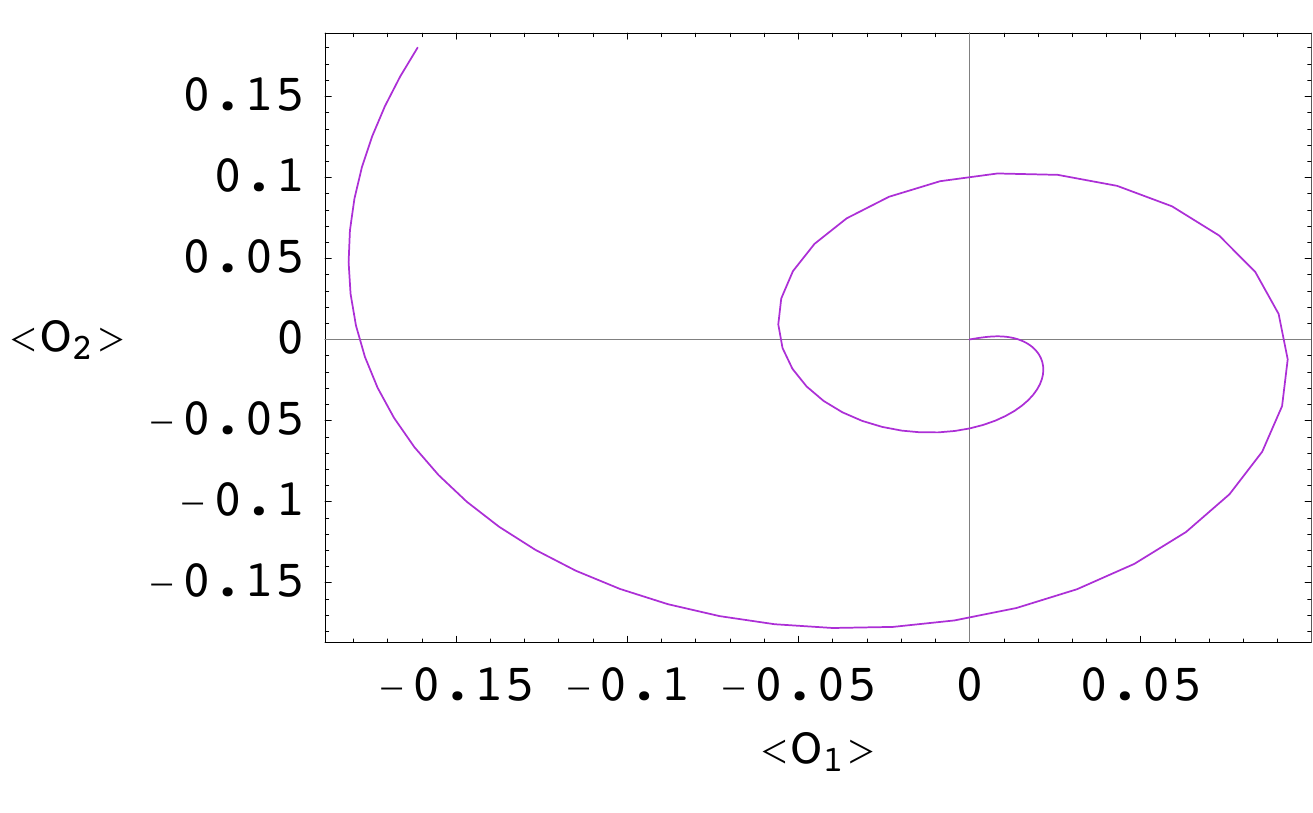} \hspace{.2cm}
\includegraphics[scale=0.5]{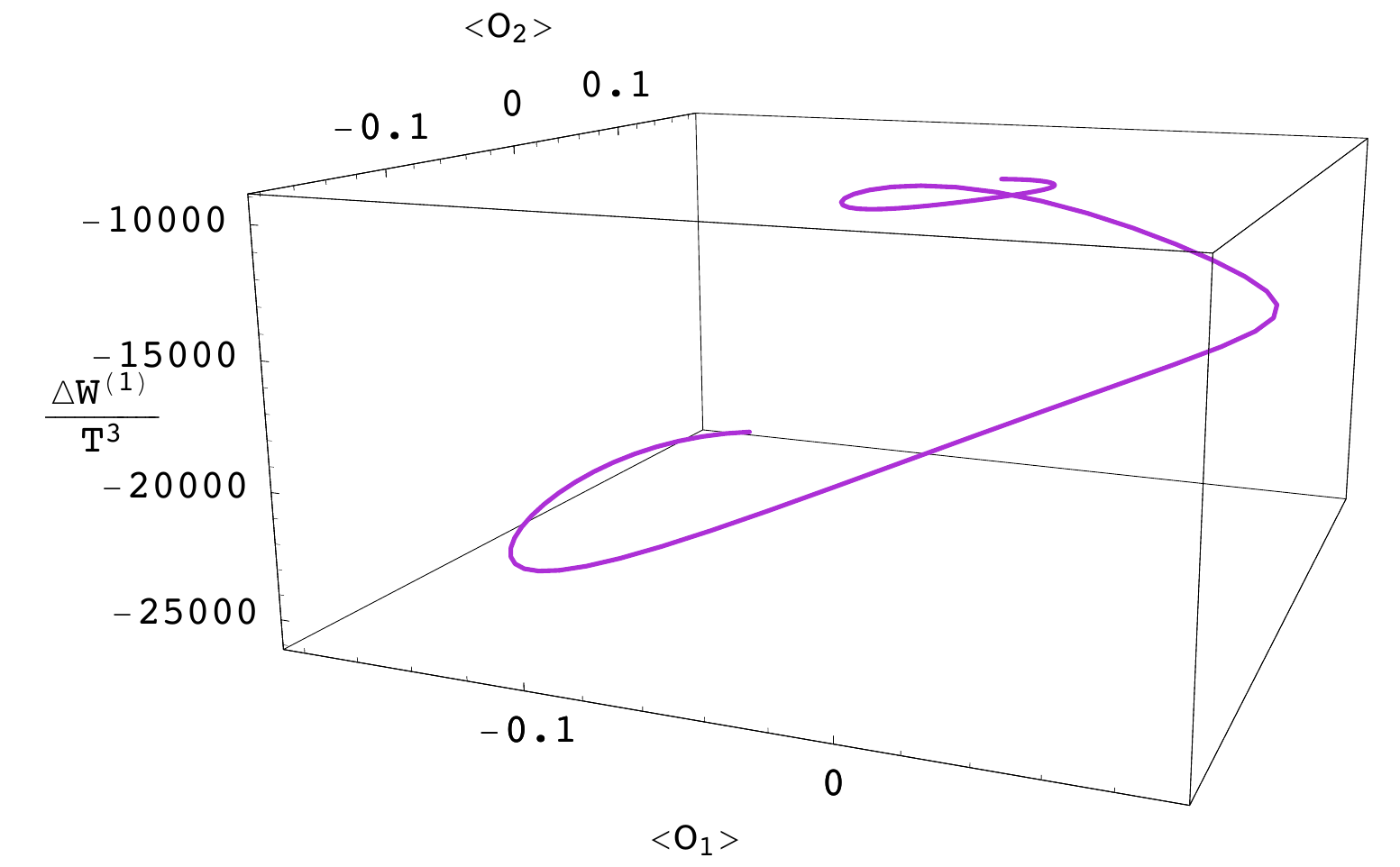} \\
\caption{Normalized $\Delta W^{(1)}$ as a function of $(\mathcal{O}_1,\mathcal{O}_2)$ for $n=2$, $\mu=-1$ and $T=0.015$.}
\label{spiral_T0015}
\end{figure}

We are now ready to compute the free energy as a function of temperature by 
evaluating (\ref{W1}) with fields given by the solution to the equations of motion. An interesting quantity is the difference in free energies between the condensed and uncondensed phases,
\begin{equation} 
 \Delta W^{(1)}=W^{(1)}_{Condensed}-W^{(1)}_{Uncondensed}.
\end{equation}
where $W^{(1)}_{Uncondensed}$ is the free energy for $\Psi_1=\Psi_2= 0$.
To get a flavor of the behavior of the free energy, we can extend our previous plots by adding it as third dimension to the spirals in the $(\mathcal{O}_1,\mathcal{O}_2)$ plane, as shown in figure \ref{spiral_T0015}. The free energy typically increases as we approach the origin, i.e. the uncondensed phase. A more thorough numerical analysis shows that the external branch (in the case of $\Delta W^{(1)}$, the branch associated with the outermost cut between the spiral and the $\mathcal{O}_2=0$ axis) is a local minimum of the free energy. On the other hand, internal branches are critical points but not extrema. In this way, we confirm the previous statement that only the external branch corresponds to a condensed phase. The local minimum becomes more manifest as we approach the critical temperature. 

\begin{figure}[h!]
\centering
\includegraphics[scale=0.7]{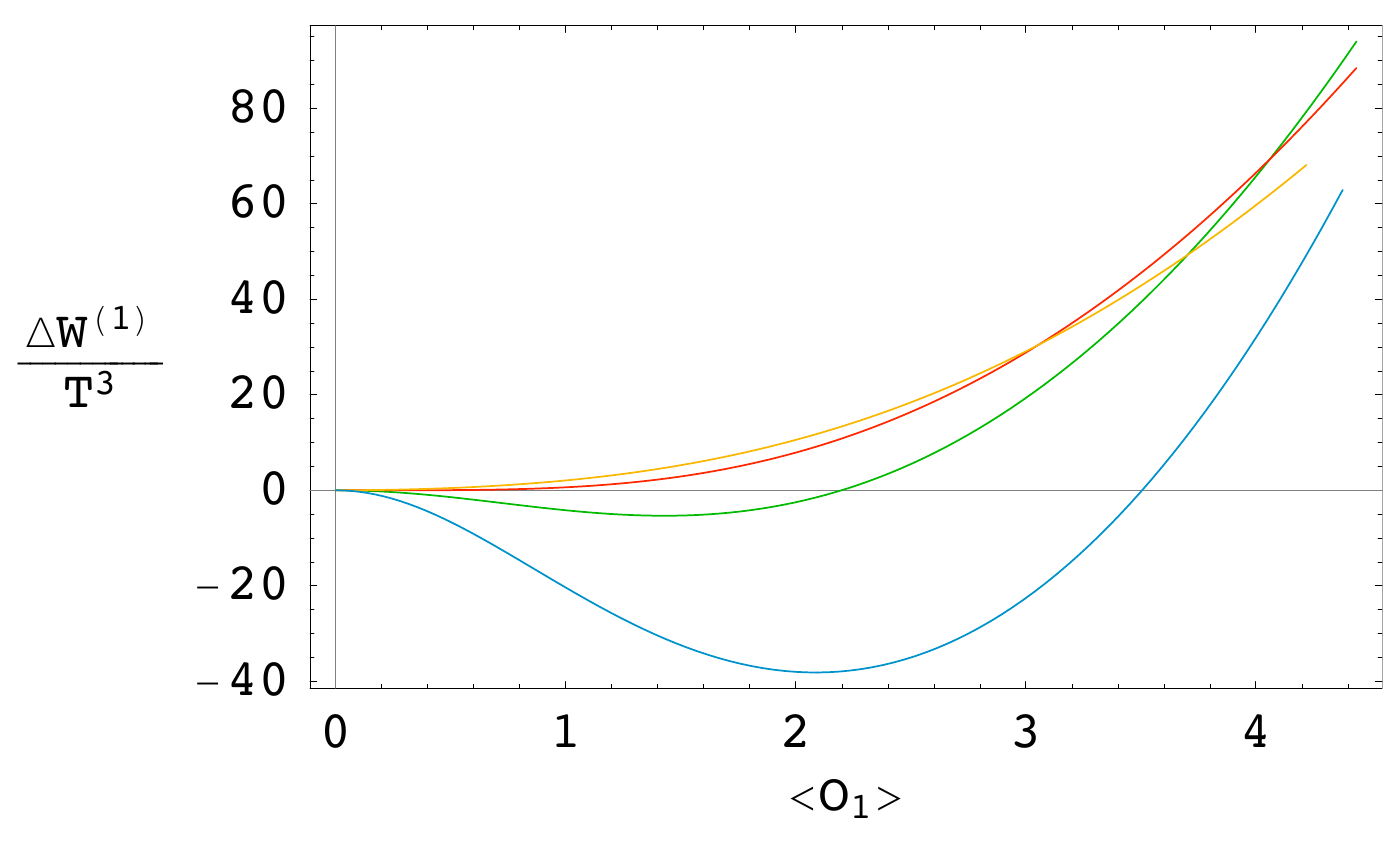}\caption{Normalized $\Delta W^{(1)}$ as a function of $\langle \mathcal{O}_1 \rangle$ for $n=2$, $\mu=-1$ and $T=0.20,0.25,0.30,0.35$.}
\label{W_transition_n2}
\end{figure}

Figure \ref{W_transition_n2} shows the free energy as a function of the condensate around the critical temperature. We observe the characteristic behavior of a second order phase transition. Above $T_c$, the only minimum is at the origin. Below $T_c$, the origin becomes a maximum and a minimum develops at non-zero condensate, which smoothly approaches zero as $T\to T_c$. The minimum in the $\mathcal{O}_1$ direction occurs at $\langle \mathcal{O}_2\rangle=0$.

\begin{figure}[h!]
\centering
\includegraphics[scale=0.7]{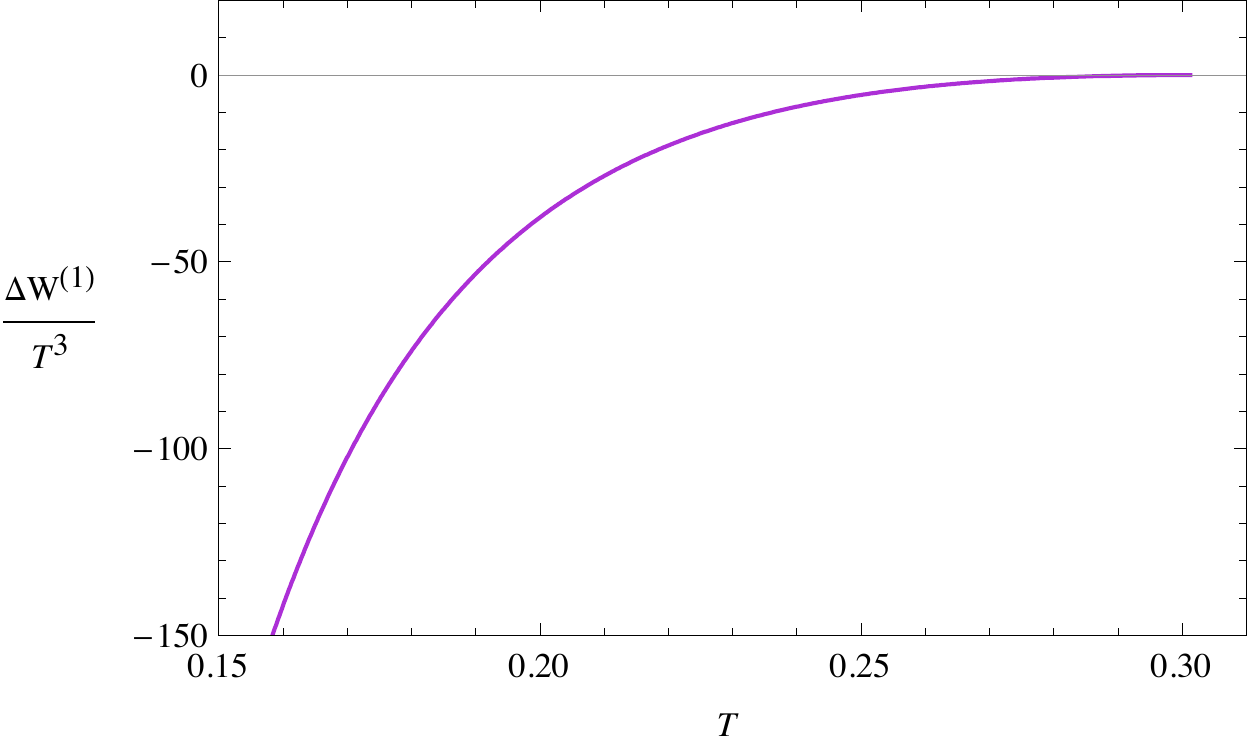}\caption{Normalized difference in free energy between the uncondensed and condensed phases as a function of the temperature.}
\label{WO1n2atfixedmu}
\end{figure}

%
Finally, figure \ref{WO1n2atfixedmu} shows the smooth convergence of the free energies of the condensed and uncondensed phases as the temperature approaches $T_c$

\subsection{Free energy for the $n=3$ model}

\begin{figure}[h!]
\centering
\includegraphics[scale=0.7]{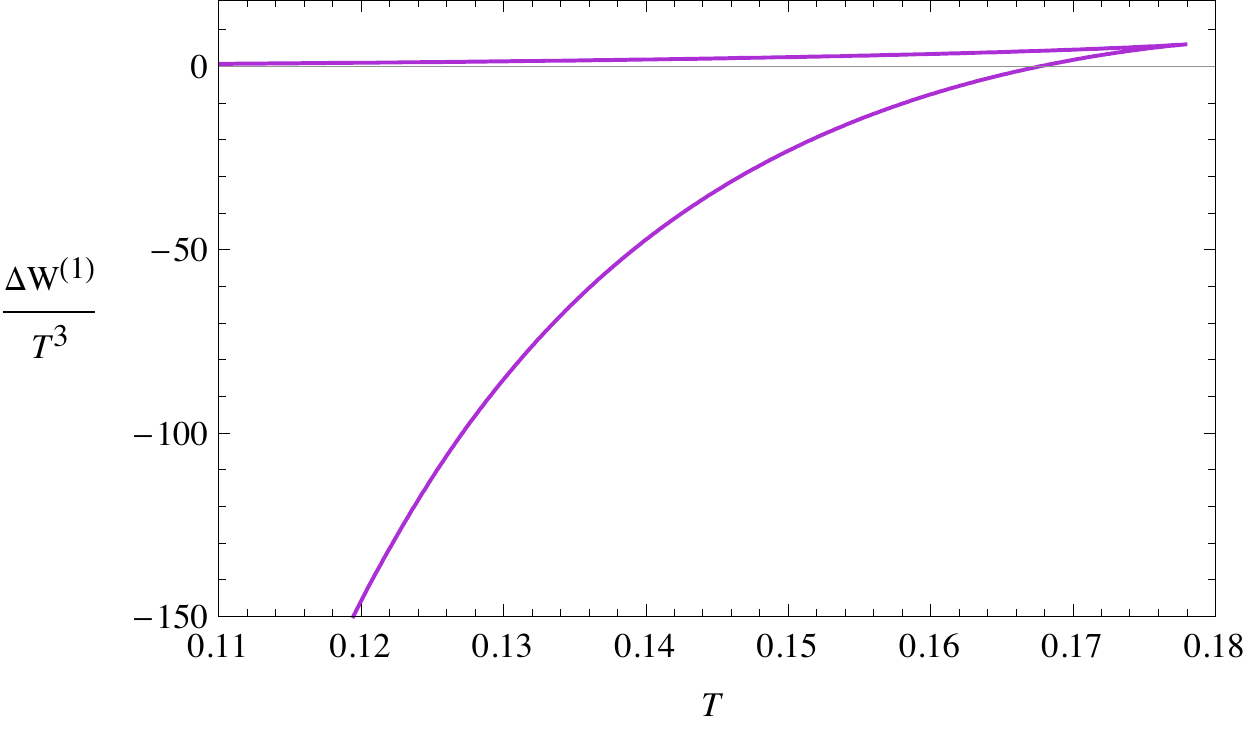}\caption{Normalized $\Delta W^{(1)}$ for the two branches as a function of the temperature.}
\label{WO1n3atfixedmu}
\end{figure}

We can use our results on the free energy to get a better understanding of the phase transition in the 
$n=3$ case. An important difference between $n=3$ and $n=2$ is the existence of a lower branch in figure \ref{O1fullCanonicalEnsemble}. Figure \ref{WO1n3atfixedmu} shows the free energy for both branches. The branch with the largest condensate has less free energy. Furthermore, it is a local minimum of the free energy, while the other branch is a local maximum. We conclude that, as we previously assumed, the system prefers to condense in the branch with the largest condensate.

\begin{figure}[h!]
\centering
\includegraphics[scale=0.7]{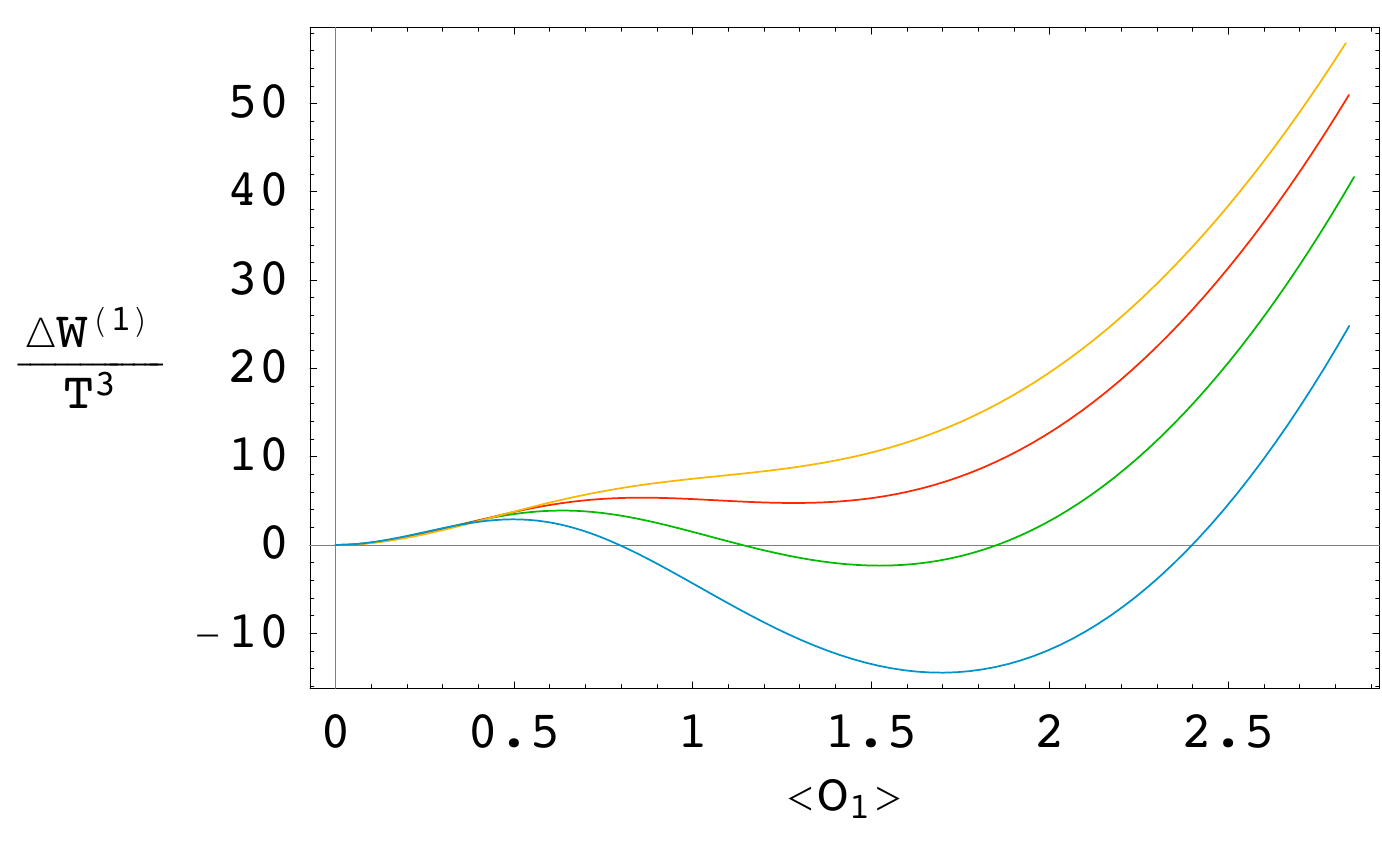}\caption{Normalized $\Delta W^{(1)}$ as a function of $\mathcal{O}_1$ for $n=3$, $\mu=-1$ and $T=0.155,0.165,0.175,0.185$.}
\label{W_transition_n3}
\end{figure}

Figure \ref{W_transition_n3} shows the free energy as a function of the condensate for various temperatures. Below some temperature $T_0$, a metastable
minimum develops at a finite value of the condensate. As we lower the temperature, we reach the critical temperature $T_c$, at which 
this minimum becomes degenerate with the uncondensed one. Since, at $T_c$, this minimum is at a finite distance from the origin, the transition
is first order. For $T<T_c$, the system is in the condensed phase.

\section{A first look at transport properties}

One of the most interesting applications of holographic techniques to strongly coupled field theories is the computation of transport properties. In the presence of an external, time-dependent electric field of the form $A_x \sim e^{-i\omega t}$, the conductivity reads (see e.g \cite{Hartnoll:2009sz,Herzog:2009xv})

\begin{equation}
\sigma(\omega)=\frac{G_R(\omega,0)}{i\, \omega}.
\end{equation}
We can use the holographic dual to compute the Fourier transform of the retarded Green's function for our CFT$_{d-1}$ at zero spatial momentum. Following the usual AdS/CFT dictionary, this boils down to computing the fluctuation of the now time dependent $A_x$ component of the bulk gauge field in the condensed background (i.e with a non-trivial profile of $\Psi$). The boundary behavior of such a fluctuation is

\begin{equation}
A_x\sim\ A_x^{(0)}+\frac{A_x^{(1)}}{r}+\cdots .
\end{equation}
From here, we can determine the Green's function, which results in the conductivity

\begin{equation}
\sigma(\omega)=-\frac{i\, A_x^{(1)}}{\omega\, A_x^{(0)}}.
\end{equation}
In summary, the main task is to solve the equations of motion for the $A_x$ fluctuation in the condensed background, 
\begin{equation}
A_x''+\frac{f'}{f}\, A_x'+\Big(\frac{\omega^2}{f^2}-\frac{2\,\Psi^n}{f}\Big)\, A_x=0,
\end{equation}
where the prime denotes derivative with respect to $r$.

In addition, we have to impose causal boundary conditions, i.e. incoming boundary conditions at the horizon. This requires that, close to the horizon,

\begin{equation}
A_x\sim (1-z)^{-\frac{i\, \omega\, L^2}{3\, r_H}}.
\end{equation}

\begin{figure}[h!]
\begin{center}
\includegraphics[scale=0.3]{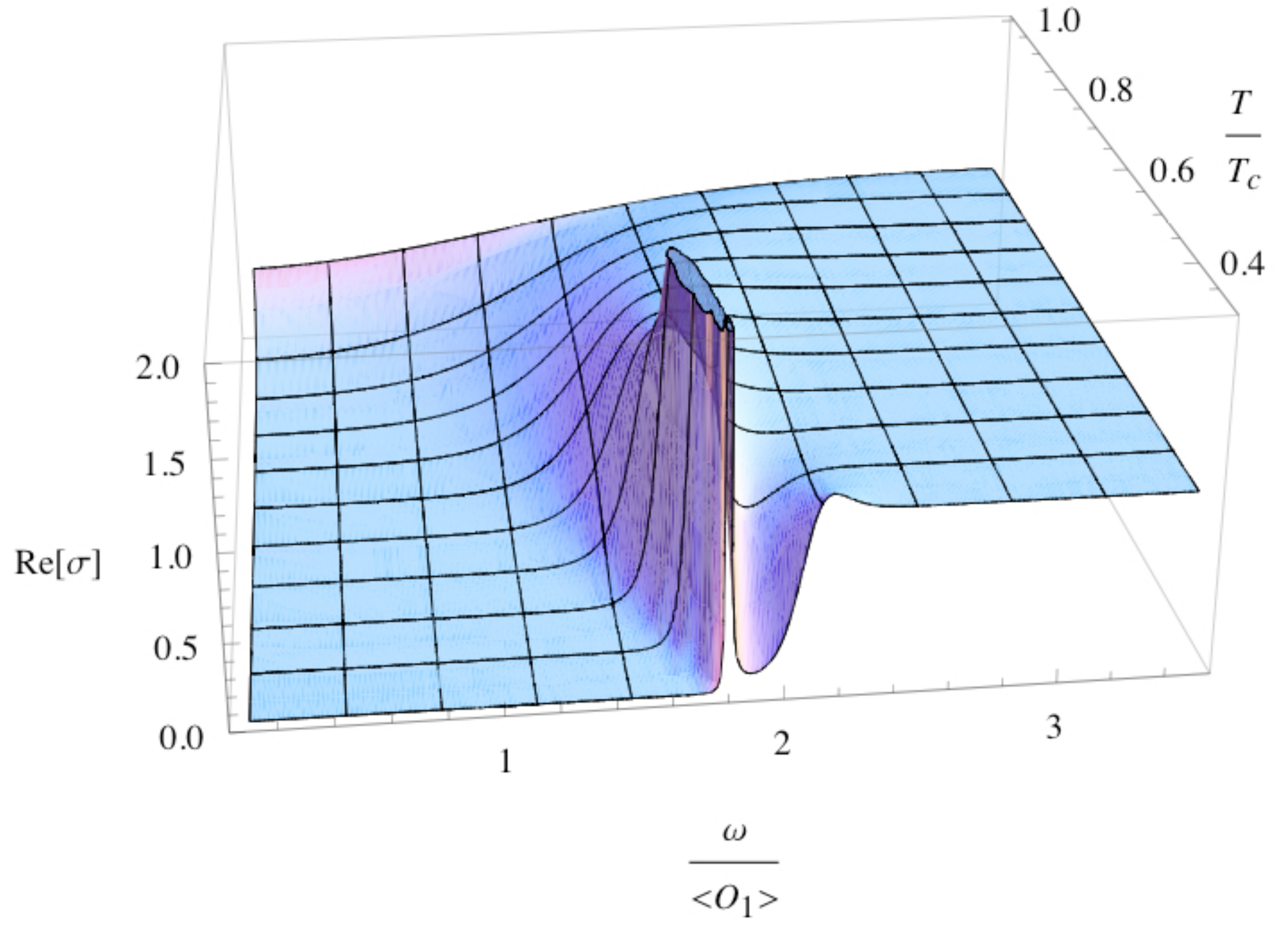}
\hspace{0.1cm}
\includegraphics[scale=0.3]{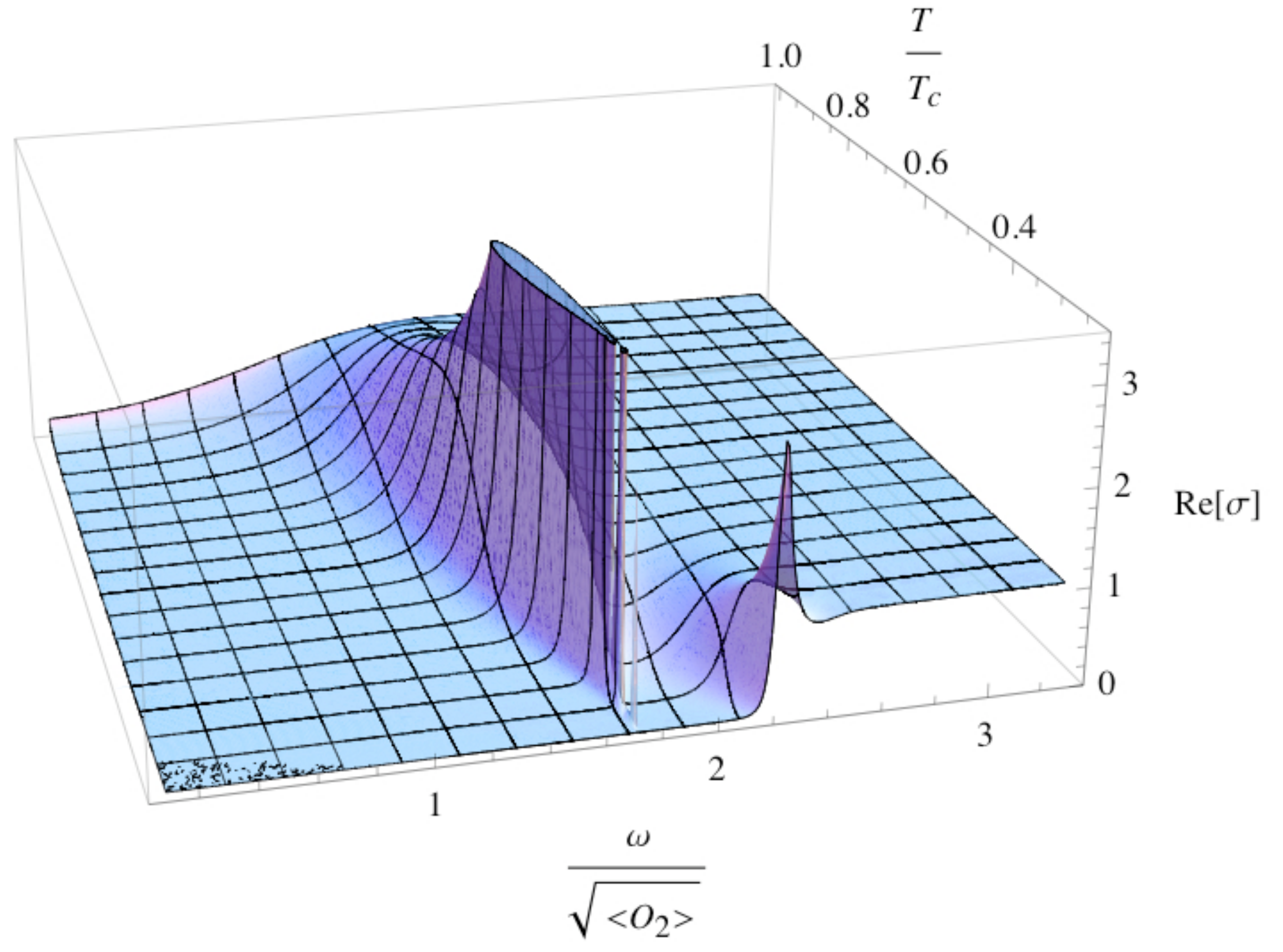}
\end{center}
\caption{${\rm Re}[\sigma]$ as a function of $\omega$ and $T$.}
\label{conductivity}
\end{figure}

The conductivity for $n=2$ was worked out in great detail in \cite{Hartnoll:2008vx} so we will focus on $n=3$. In figure \ref{conductivity}, we plot the real part of the conductivity as a function of $\omega$ for both $\mathcal{O}_1$ and $\mathcal{O}_2$ condensates at various temperatures. 
 
Figure \ref{conductivity} clearly shows the existence of a gap. The most interesting difference with respect to $n=2$ is the appearance of resonances at certain non zero frequencies. These additional poles suggest the existence of bound states caused by strongly interacting low energy excitations (quasiparticles) of this superconductor. 
This behavior is qualitatively similar to the one observed in \cite{Horowitz:2008bn} for holographic superconductors in various dimensions as the bulk mass of the scalar approaches the BF bound. It would be interesting to analyze the transport properties of this model in further detail.

\section{More general models}

In this section we explore models with more general forms of the function $\mathcal{F}$. For this purpose, let us focus on 

\begin{equation}
\mathcal{F}=\Psi^2 + c_3 \Psi^3 + c_4 \Psi^4.
\label{F_2-3-4}
\end{equation}
In order to avoid keeping track of the absolute value in (\ref{gmodel}), we will consider $c_4 > 0$ and values of $c_3$ such that $\mathcal{F} \ge 0$ for all $\Psi$.

\begin{figure}[h!]
\centering
\includegraphics[scale=0.6]{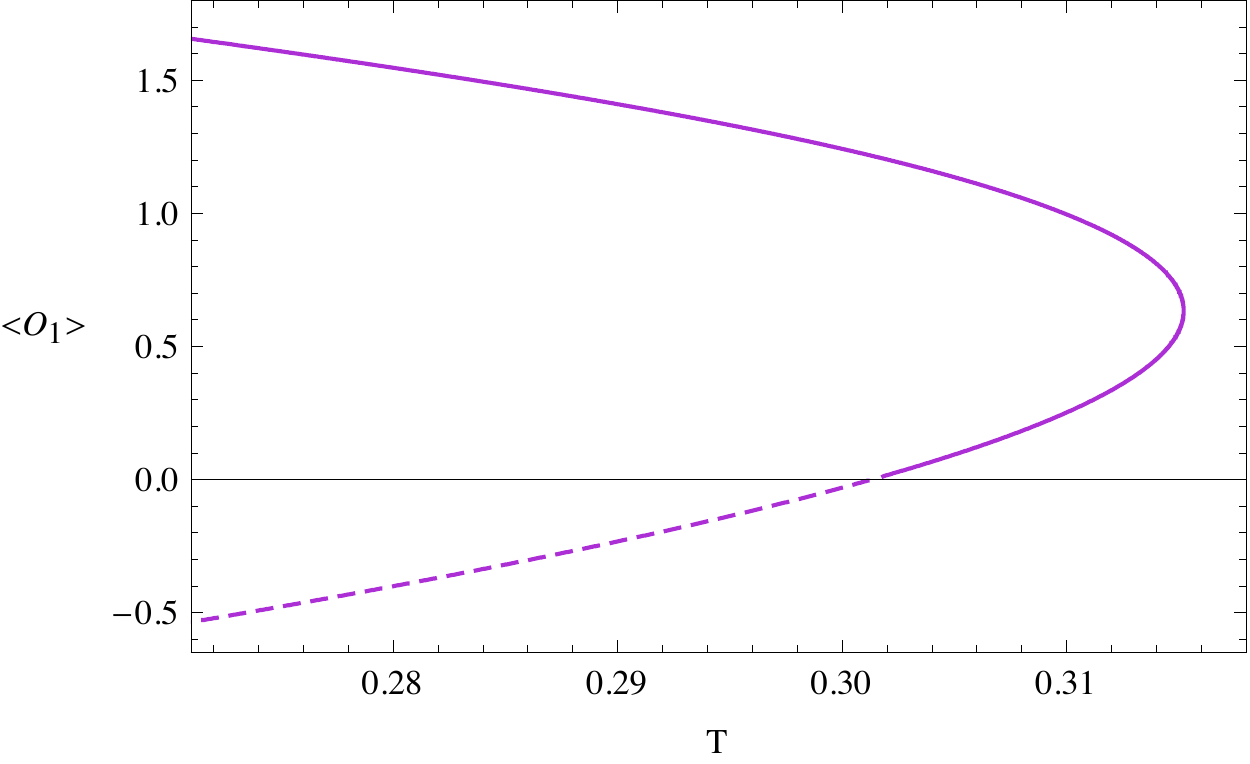} \hspace{.2cm}
\includegraphics[scale=0.6]{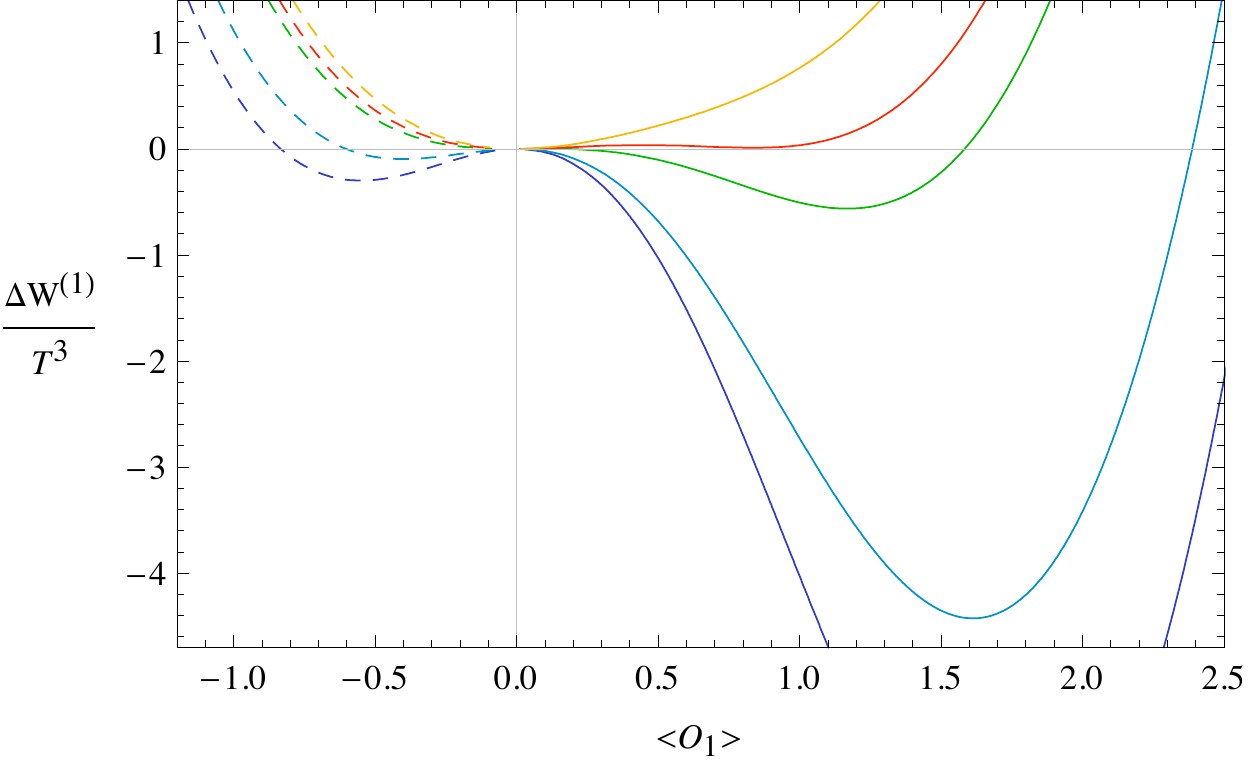} \\
\vspace{-.1cm}\hspace{1cm}{\footnotesize $(a)$} \hspace{7.3cm} {\footnotesize $(b)$} 
\caption{(a) $\mathcal{O}_1$ condensate as a function of temperature at fixed $\mu=-1$ for $c_3=-1$ and $c_4=1/4$. (b) Normalized $\Delta W^{(1)}$ at fixed $\mu=-1$ as a function of the condensate for $T=0.270$, $0.280$, $0.305$, $0.314$ and $0.330$.}
\label{positive_c3}
\end{figure}

Figure \ref{positive_c3}.a shows the condensate as a function of temperature for $c_3,c_4>0$. The free energy is plotted in Figure \ref{positive_c3}.b. We conclude there is a first order phase transition with a positive condensate. The non-vanishing $c_3$ breaks the $\Psi \to - \Psi$ symmetry of previous examples. As we discussed around (\ref{gmodel}), it is sometimes possible to constraint $\Psi \ge 0$ by deriving the model from one with a complex field. In this case, only the solid pieces of the curves in Figure \ref{positive_c3}. Complex models with the desired properties indeed exist for (\ref{F_2-3-4}); we do not present them here for simplicity. 

\begin{figure}[h!]
\centering
\includegraphics[scale=0.6]{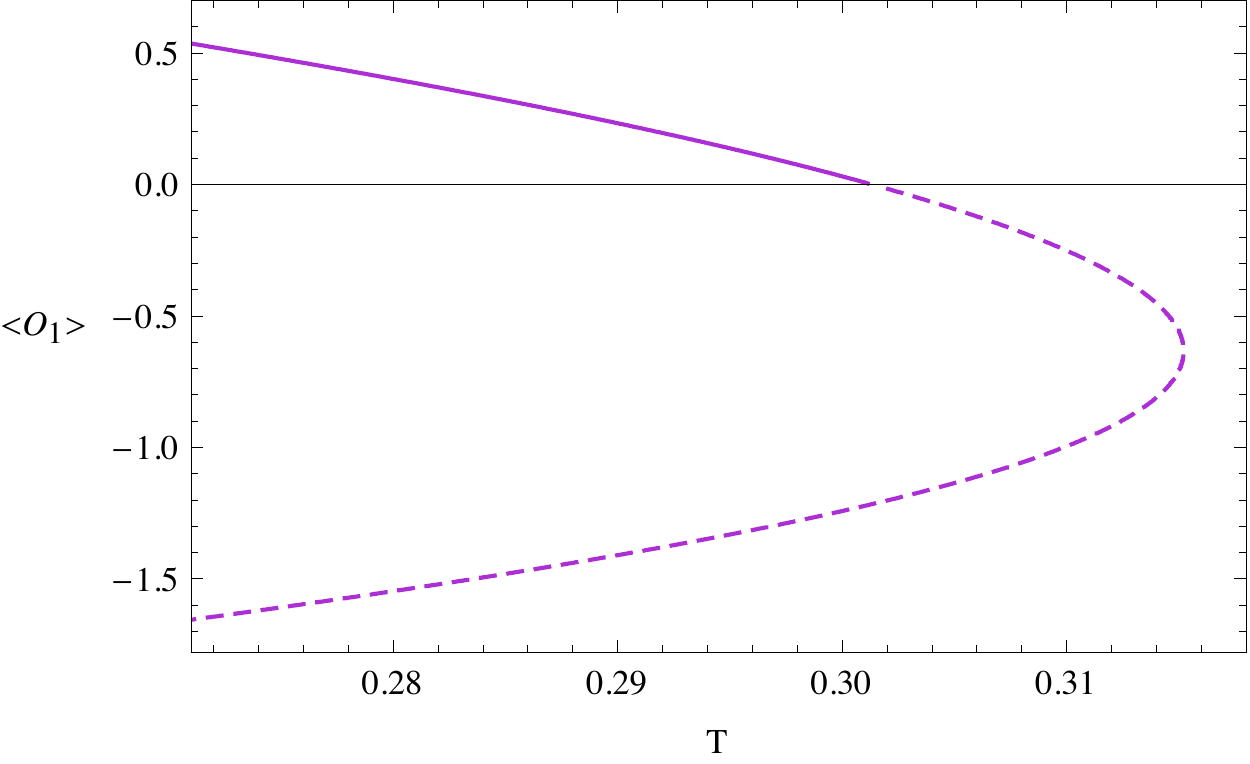} \hspace{.2cm}
\includegraphics[scale=0.6]{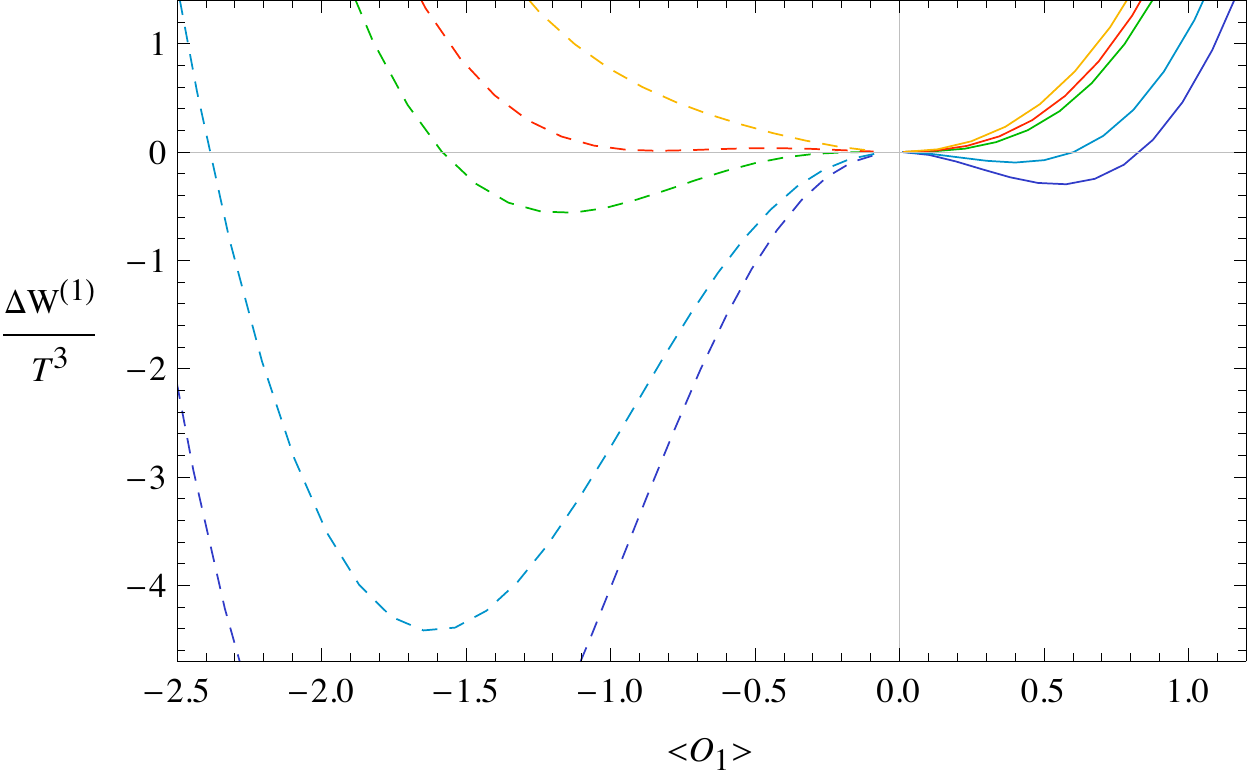} \\
\vspace{-.1cm}\hspace{1cm}{\footnotesize $(a)$} \hspace{7.3cm} {\footnotesize $(b)$} 
\caption{(a) $\mathcal{O}_1$ condensate as a function of temperature at fixed $\mu=-1$ for $c_3=-1$ and $c_4=1/4$. (b) Normalized $\Delta W^{(1)}$ at fixed $\mu=-1$ as a function of the condensate for $T=0.270$, $0.280$, $0.305$, $0.314$ and $0.330$.}
\label{negative_c3}
\end{figure}

Let us now consider negative $c_3$. Figure \ref{negative_c3}.a is similar to the corresponding one for positive $c_3$, but shifted towards negative $\mathcal{O}_1$. Figure \ref{negative_c3}.b shows the free energy. If $\Psi$ can take any real value, we conclude there is a first order phase transition with a negative condensate. As before, if $\Psi\ge 0$ only the solid branches survive. Interestingly, in this case the transition is second order.\footnote{Notice that $O_1=0$ is a critical point of the free energy for both positive and negative $c_3$, with no apparent non-analyticity when keeping the positive branch.} 

In the case of second order phase transitions, it is of great interest to determine the critical exponent $\beta$ that controls the vanishing of the condensate as the temperature approaches $T_c$

\begin{equation}
\mathcal{O}_i(T) \sim A \, (T_c - T)^\beta.
\end{equation}
The $n=2$ model was analyzed in \cite{Hartnoll:2008kx}, where it was found that 
$\beta=1/2$ for both $\mathcal{O}_1$ and $\mathcal{O}_2$.\footnote{Other critical exponents of this model were studied in \cite{Maeda:2008ir,Maeda:2009wv}.} This value of the critical exponent is the same one that appears in the mean field approximation of Landau-Ginzburg. In the present case, the condensate vanishes as

\begin{equation}
\mathcal{O}_1(T)\sim \mathcal{O}_1'(T_c) \, (T_c-T), 
\end{equation}
since ${O}_1'(T_c)$, the derivative of the condensate with respect to the temperature evaluated at $T_c$, is finite. Namely, this corresponds to $\beta=1$. This behavior just follows from vertically shifting the condensate curve and restricting it to the positive branch. It would be interesting to investigate whether it is possible to find more tunable ways of obtaining non mean-field critical exponents.

\section{Conclusions}

We have introduced a general class of strongly coupled CFT$_{d-1}$'s with dual holographic descriptions which exhibit spontaneous symmetry breaking of a global $U(1)$ symmetry at low temperatures. This class contains and generalizes the model in \cite{Hartnoll:2008vx}. Our dual models are defined in terms of a St\"uckelberg-like lagrangian coupled to gravity which is determined by a function $\mathcal{F}$. We considered some basic forms of  $\mathcal{F}$, given by order $n$ monomials and combinations of them. 

Interestingly, for $n\ge 3$ the phase transition is of first order, while for $n=2$ it is of second order. We have explicitly seen this by studying the free energy, whose derivative exhibits the characteristic discontinuity for $n\ge 3$. Interestingly, metastability, one of the characteristic features of a first order phase transition, is also clearly observed. 

We have initiated the study of the transport properties of these systems. The most striking feature is the appearance of extra resonances. Perhaps this feature is not so surprising, since for $n\ge 3$ we can understand the ``vertex" $\Psi^n(\partial p-A)^2$ along the lines of \cite{Polchinski:2002jw} as leading to inelastic scattering. Therefore, it is plausible that for $n\ge 3$ the condensate, represented by its dual field $\Psi$, does actually have some internal structure which in particular allows its breaking. This would show up in the conductivity as extra poles. It would be interesting to understand the dependence of the gap on $n$ for monomial $\mathcal{F}$ or a more detailed calculation of the conductivity in order to determine whether it reaches some limiting curve.

We have taken a step towards the study of general $\mathcal{F}$, by considering the model in (\ref{F_2-3-4}). Interestingly, when the condensate is constrained to be positive because the model comes from one with a complex scalar, it has a second order phase transition with a critical exponent $\beta=1$. This behavior is different from the $1/2$ exponent characteristic of the mean field approximation that arises for the pure $n=2$ model. It would certainly be interesting to study whether, either using models with generalized kinetic terms as the ones introduced in this paper or by other means, it is possible to obtain other critical exponents. Finally, one of the most exciting remaining questions is how the details of the function $\mathcal{F}$ are mapped to the dual CFT.

 \section*{Acknowledgments}

We are grateful to G. Horowitz, T. Ort\'{\i}n, M. Roberts and A. Yarom for useful discussions. We wish to specially thank Chris Herzog for innumerable and patient explanations. D.R-G would like to thank Yolanda Lozano for the organization of the \textit{First Iberian Meeting} held in Gij\'on (Spain) where a substantial part of this work was done. S. F. is supported by the National Science Foundation under Grant No. PHY05-51164. D. R-G. acknowledges financial support from the European Commission through Marie Curie OIF grant contract No. MOIF-CT-2006-38381. A.M. G. acknowledges financial support from both a Marie Curie Outgoing Action, contract MOIF-CT-2005-007300 and the FEDER
and the Spanish DGI for financial support through Project No.
FIS2007-62238.


\end{document}